\documentclass[10pt]{article}
\usepackage{bm,color}
\usepackage{hyperref}
\usepackage{amsfonts,epsfig}
\newcommand{\orb}{{\sc Orb5}}
\newcommand{\dd}{{\rm d}}
\newcommand{\dt}{{\rm dt}}

\newcommand{\bfZ}{{\mathbf{Z}}}

\begin{document}
\title{First principles gyrokinetic analysis of electromagnetic plasma instabilities}
 \maketitle
\author{N Tronko$^{1,2}$, A Bottino$^1$, C Chandre$^3$, E Sonnendr\"ucker$^{1,2}$,\\ E Lanti$^4$, N Ohana$^4$, S Brunner$^4$,  and L Villard$^{4}$}
\\
$^1$ Max-Planck Institute for Plasma Physics, 85748, Garching, Germany\\
$^2$ TU Munich, Mathematics Center, 85747, Garching, Germany\\
 $^3$ Aix Marseille Univ, CNRS, Centrale Marseille, I2M, Marseille, France\\
 $^4$ Swiss Plasma Center, Ecole Polytechnique F\'ed\'erale de Lausanne, CH-1015, Lausanne, Switzerland
\date{\today}
%\ead{nataliat@ipp.mpg.de}
%\vspace{10pt}
%\begin{indented}
%\item[]March 1, 2019
%\end{indented}

\begin{abstract}
A two-fold analysis of electromagnetic core tokamak instabilities in the framework of the gyrokinetic theory is presented.  First principle theoretical foundations of the gyrokinetic theory are used to explain and justify the numerical results obtained with the global electromagnetic particle-in-cell code {\orb}  whose model is derived from the Lagrangian formalism.
The energy conservation law corresponding to the {\orb}    model is derived from the Noether theorem and implemented in the code as a diagnostics for energy balance and conservation verification.
An additional Noether theorem based diagnostics is implemented in order to analyse destabilising mechanisms for the electrostatic and the electromagnetic Ion Temperature Gradient (ITG) instabilities in the core region of the tokamak. The transition towards the Kinetic Ballooning Modes (KBM) at high electromagnetic $\beta$ is also investigated.
\end{abstract}

\section{Introduction}
Strongly magnetised fusion plasmas represent a paradigmatic example of out-of-equilibrium systems, in which turbulence is ubiquitous. This omnipresence originates from the concept of magnetic fusion itself: Bringing the mix of hydrogen isotopes into the confinement mode implies by construction the existence of strong spatial gradients. Typically, a difference of three orders of magnitude for the temperature is present, with $\sim10^8 K$ at the centre of the device, where the plasma is hot and relatively dense and with $\sim 10^5 K$ close to the edge, where the plasma is more rarefied and colder. The strong gradients or space inhomogeneities of the temperature, velocity and current are intrinsic to fusion plasmas, and provide sources of free energy and therefore represent sources for instabilities. The instabilities manifest themselves via the exponentially growing perturbations of the electromagnetic fields. When the critical amount of instabilities has been developed, the system moves to a turbulent state with strongly unpredictable behaviour in space and time. In turn, the transport associated with turbulence is extremely dangerous for the plasma confinement. The understanding of its origins is a subject of numerous investigations \cite{Wesson}.

Extensive studies in the framework of different formalisms, from the single fluid MHD model and the multi-fluid approaches to kinetic models, aiming to identify the instability mechanisms, have been carried out over the past decades, both analytically and numerically. A detailed overview of these studies is presented in \cite{Horton_book}, where the transition between the electrostatic and the electromagnetic regimes is also discussed. In particular, in low $\beta= p/[B^2/(8\pi)]$ plasmas, where $p$ is the kinetic pressure and $B^2/(8\pi)$ the magnetic pressure, the ion temperature gradient instability, known as the ITG mode, has almost an electrostatic polarization, and therefore carries almost only electrostatic energy for the coupled drift waves and the ion acoustic waves that gives rise to a collective instability. 
 
In higher pressure plasmas with $\beta > m_e/m_i$, the electromagnetic energy is injected, and the ITG mode couples with the shear Alfv\'en wave becoming a dispersive oscillation with an electromagnetic polarization. As the plasma pressure increases the inductive electric field from the fluctuating magnetic field $\delta B_{\perp}$ begins to cancel part of the electrostatic component of the parallel electric field. This cancellation reduces the energy transfer rate $\left\langle j_{\|}E_{\|}\right\rangle$ and reduces the growth rate of instability. 
 
There are numerous studies of these weakly electromagnetic ITG modes including \cite{Rewoldt_1998}, \cite{Hong_a_1989}, \cite{Hong_b_1989} and \cite{Kim_1993} which detail how the ITG modes change with increasing $\beta$. At certain intermediate values of $\beta$, both the ITG and the kinetic ballooning modes are present with different frequencies and growth rates.

In this paper, we investigate the transition between the low $\beta$ and finite $\beta$ regimes in the framework of the gyrokinetic theory, both analytically and numerically.

Strongly magnetized plasma exhibits multi-scaled dynamics: The fast rotation around magnetic field lines, called gyromotion, is at least three orders of magnitude faster than the slow drifts across the magnetic field lines. The gyrokinetic dynamical reduction \cite{Brizard_Hahm},\cite{Tronko_Chandre_2018} aims to simplify the dynamical description in which fast gyration is systematically and reversibly eliminated, resulting in considerable simplifications and a gain of computational time.

Nowadays, the gyrokinetic (GK) codes play a significant role in the understanding of the development and the saturation of turbulence and the prediction of the subsequent transport properties \cite{Garbet_Idomura_2010}.
Electrostatic gyrokinetic simulations have been the topic of numerous studies during the last decades \cite{Jolliet_2007}, \cite{Jenko_2000}, \cite{Bottino_Sonnendruecker}, so that the properties of the electrostatic instabilities in the framework of the gyrokinetic theory are rather well known. However, global electromagnetic simulations are more recent \cite{Goerler_2011}, \cite{Lanti_2018}, \cite{Goerler_Tronko_2016} and some elements related to the GK electromagnetic instability mechanisms still need to be clarified.

In order to provide a better understanding of the global electromagnetic GK simulations, we present an analysis of the instability mechanisms by performing global linear electromagnetic simulations with the particle-in-cell code \orb. The numerical set-up is similar to the one used for benchmarking the global electromagnetic codes \cite{Goerler_Tronko_2016}. The \orb{} code is based on the GK Lagrangian model \cite{TBS_2016}, \cite{Lanti_2018} and allows one to use diagnostics tools issued from first principles, exactly corresponding to the theoretical model. 

This article is organised as follows: In Sec.~\ref{sec:Action_ORB5}, the field Lagrangian model including all approximations implemented in the code \orb\ is presented. Section~\ref{sec:Noether_ORB5} provides the derivation of the \orb\ energy invariant through the Noether theorem. In Sec.~\ref{sec:PB_diag}, the diagnostics issued from Noether's method are derived, and their implementation in the code \orb{} is discussed. Finally, in Sec.~\ref{sec:num_analysis}, the analysis of the electromagnetic instability mechanisms is presented.

%%%%%%%%%%%%%%%%%
\section{\label{sec:Action_ORB5}Variational formulation of the \orb\  code model}
%%%%%%%%%%%%%%%%

A detailed derivation of the \orb{} model in the framework of the Eulerian variational principle \cite{brizard_prl_2000} is given in \cite{TBS_2016}, where CGS units are used.
In this work, we use the Lagrangian variational formulation accordingly to \cite{Sugama_2000}, since the Lagrangian formulation of the gyrokinetic field theory lends itself to a discretisation by finite element methods. We notice that the choice of formalism, Eulerian or Lagrangian, does not affect the final expressions for the gyrokinetic Maxwell-Vlasov equations.

The Lagrangian action functional for the code \orb{}  depends on the fluctuating electromagnetic potentials $(\phi_1, A_{1\|})$ and on the coordinates of the Lagrangian particle trajectories $\mathbf{Z}(\mathbf{z}_0,t)=\left[\mathbf{X}(\mathbf{z}_0,t),p_z(\mathbf{z}_0,t),\mu(\mathbf{z}_0,t),\theta(\mathbf{z}_0,t)\right]$, where ${\bf X}$ are the positions of the gyrocentres, $p_z$ their momenta parallel to the magnetic field lines, $\mu$ their magnetic moments and $\theta$ their gyroangles. These coordinates are labelled by their initial conditions $\mathbf{z}_0$ such that ${\bf Z}({\bf z}_0,0)={\bf z}_0$. The expression of the action is given by \cite{TBS_2016}:
\begin{eqnarray}
{\mathcal A}^{\mbox{\scriptsize \orb} }&=&\int_{t_0}^{t_1} \dt\ 
\mathcal{L}[\mathbf{Z},\phi_1,A_{1\|}]=\sum_{\mathrm{s}} \int_{t_0}^{t_1} \dd t\  \int \dd \mathbf{z}_0\ f_{0,s}(\mathbf{z}_0)\ L_s \left(\mathbf{Z}(\mathbf{z}_0;t),\dot{\mathbf Z}(\mathbf{z}_0;t)\right)
\nonumber
\\
&-&\epsilon_{\delta}^2\sum_{\mathrm{s}} \int_{t_0}^{t_1} \dd t\ \int \dd {\mathbf z}_0\ f_{C,s} (\mathbf Z(\mathbf{z}_0;t))
 \ H_{2,s}(\mathbf Z(\mathbf{z}_0;t))  \nonumber \\
 && -\epsilon_{\delta}^2\int_{t_0}^{t_1} \dd t\  \int \dd \mathbf{x} \ \frac{\left| {\bm\nabla}_{\perp} { A}_{1\|}\right|^2}{8\pi},
\label{L_ORB5}
\end{eqnarray}
where the particle Lagrangian is given by:
 \begin{equation}
 L_s(\mathbf{Z},\dot{\mathbf Z})=\frac{q_s}{c} \mathbf{A}_s^*\cdot\dot{\mathbf{X}}+\frac{m_s c}{q_s}\mu\dot{\theta}-\left(H_{0,s}+\epsilon_{\delta} H_{1,s}\right),
 \end{equation}
the generalized vector potential is defined as:
\begin{equation}
\label{eq:A*}
{\mathbf A}_s^*={\mathbf A}+\frac{c}{q_s}\ p_z\ \widehat{\mathbf b}.
\end{equation}
The volume element in the reduced phase space is $\dd \mathbf{z}_0= m B_{\|,s}^*({\bf z}_0)\dd\mathbf{X}_0\dd p_{z,0} \dd\mu_0 \dd\theta_0$ (see Ref.~\cite{Cary_Brizard} for more detail) where 
\begin{equation}
\label{eq:B*}
\mathbf{B}_s^*=\bm\nabla\times{\mathbf A}_s^* \qquad \mbox{and } B_{\| , s}^*=\widehat{\mathbf{b}}\cdot\mathbf{B}_s^*.
\end{equation}
The dynamical distribution function is denoted by $f_s$ for each species, and the background distribution function by $f_{C,s}$. The property of the Vlasov distribution function being conserved along the particles trajectories translates into
\begin{equation}
\label{Euler_Largange}
f_{0,s}(\mathbf{z}_0)=f_s(\mathbf{Z}(\mathbf{z}_0,t);t).
\end{equation}
  We assume that the background distribution $f_{C,s}$ is frozen and corresponds to  a good approximation of $f_s$ at all times. 
 The first and the second terms of the action~(\ref{L_ORB5}) correspond to the gyrocentre reduction, and the last term is a contribution from the
perturbed magnetic field. We notice that only the perpendicular part of the perturbed magnetic field
$\mathbf{B}_{\perp}=\widehat{\mathbf b}\times\bm\nabla A_{1\|}$ is
implemented.

The background Hamiltonian contains information on the kinetic energy of a charged particle moving in
a magnetic field of amplitude $B$:
\begin{equation}
H_{0,s}= \frac{p_z^2}{2 m_s}+\mu B.
\label{H_{0,s}}
\end{equation}
The first order correction of the Hamiltonian model for the ions is given by 
\begin{equation}
H_{1,s}=q_s\left\langle\phi_1(\mathbf{X}+{\bm\rho}_{0,s})-A_{1\|}(\mathbf{X}+{\bm\rho}_{0,s})\frac{p_{z}}{m_s c} \right\rangle,
\label{H_1}
\end{equation}
where the gyroaveraging operator $\langle \cdot \rangle$ is the average over the fast gyroangle $\theta$ contained in the fast rotating Larmor vector ${\bm\rho}_{0,s}$ measuring the distance between the initial particle position and the guiding-centre position.  More precisely, in an orthonormal basis $(\hat{\bf b}_1,\hat{\bf b}_2,\hat{\bf b})$ at ${\bf X}$, the Larmor vector is given by ${\bm\rho}_{0,s}=mc/q_s \sqrt{2\mu/m_s B} [\hat{\bf b}_1({\bf X})\cos\theta -\hat{\bf b}_2({\bf X})\sin\theta]$, and the expression of the gyroaveraging is given by
\begin{equation}
\left\langle\phi(\mathbf{X}+\bm\rho_{0,s})\right\rangle=\frac{1}{2\pi}\int_0^{2\pi}\dd \theta\ \phi(\mathbf{X}+\bm\rho_{0,s}).
\end{equation}
For the electrons,  the first order correction to the Hamiltonian only contains the first order Finite Larmor Radius (FLR) correction. It corresponds to the drift-kinetic dynamics, and it is given by
\begin{equation}
H_{1,e}= e\left( \phi_1(\mathbf{X})-A_{1\|}(\mathbf{X})\frac{p_{z}}{m_e c}\right).
\label{H_1}
\end{equation}
In what follows the bracketed quantities are evaluated at the position $\mathbf{X}+\bm{\rho}_{0,s}$, all the other quantities are evaluated at the gyrocentre position $\mathbf{X}$.

There exist several nonlinear models in the \orb{} code.
The most complete model considers the nonlinear Hamiltonian model  $H_2$ for the ions, including all order FLR corrections in its electrostatic part and up to second order FLR terms in its electromagnetic part.
In this work, in addition to the drift-kinetic model for the electrons, we consider the \textit{long-wavelength approximation} of the nonlinear model for the ions (see Eq.~(68) in Ref.~\cite{TBS_2016}), i.e. :
\begin{eqnarray}
\label{H_2FLR}
H_{2,s}=-\frac{m_s c^2}{2B^2}\left|\bm\nabla_{\perp}\phi_1\right|^2
+\frac{q_s^2 }{2 m_s c^2}\left(A_{1{\|}}^2+\left(\frac{m_s c^2}{q_s^2}\right) \frac{\mu}{B}A_{1\|}\bm\nabla_{\perp} ^2A_{1\|}\right).
\end{eqnarray}
The second order Hamiltonian model for the electrons only contains the drift-kinetic correction:
\begin{eqnarray}
\label{H_2_e}
H_{2,e}&=&\frac{e^2 }{2 m_e c^2}\ A_{1{\|}}^2.
\end{eqnarray}
Before presenting the equations of motion implemented in \orb{}, we discuss all necessary
approximations included in the gyrokinetic action given by Eq.~(\ref{L_ORB5}).  The first term
of the action involves the full distribution function $f_s$, while the second term involving the
nonlinear Hamiltonian $H_{2}$ contains a ``canonical'' distribution function $f_{C,s}$, which is by
definition invariant under the unperturbed Hamiltonian dynamics, i.e., it satisfies the
condition $\{f_{C,s},H_{0,s}\}_{\mathrm{gc}}=0$, where the guiding-center Poisson bracket is defined accordingly to Eq.~(7) in Ref.~\cite{TBS_2016}. This approximation brings several simplifications to the model. First, it
results in the linearisation of the gyrokinetic Poisson and Amp\`ere equations. Second, it simplifies the
gyrokinetic Vlasov equation by excluding some nonlinear terms from the gyrocentre characteristics
associated with Hamiltonian $H_2$.

%%%%%%
\subsection{Gyrokinetic field equation}
%%%%%%
The equations of motion are derived from $$\delta\mathcal{A}^{\mbox{\scriptsize \orb}} = \delta \int_{t_0}^{t_1} \mathcal{L} \,\mathrm{d} t =0.$$ 
We use functional derivatives for evaluating the r.h.s. of this expression explicitly.
As a reminder, for a functional $\mathcal{L}[\eta]=\int \dd^n{\bf r} \ {\mathsf L}\left(\eta,\bm\nabla\eta\right)$ depending functionally on a scalar field $\eta$ and its gradient $\bm \nabla\eta$,
the functional derivative is defined as $\delta{\mathcal L}/\delta\eta$ acting on the test function $\chi$ as:
\begin{eqnarray}
\int \dd^n{\mathbf r}\  \frac{\delta{\mathcal L}}{\delta\eta }\cdot\chi(\mathbf{r}) &=&
\left.\frac{\rm d}{{\rm d}\nu} \left[\int \dd^n{\bf r}\  {\mathsf L}\left(\eta+\nu\chi,\bm\nabla\eta+\nu\bm\nabla\chi\right)\right]\right|_{\nu=0}
\nonumber
\\
&=&
\int \dd^n{\bf r} \ \frac{\partial{\mathsf L}}{\partial\eta} \chi+\int \dd^n{\bf r}\ \frac{\partial{\mathsf L}}{\partial\bm\nabla\eta}\cdot\bm\nabla\chi.
\label{weak_eqns}
\end{eqnarray}
The corresponding quasineutrality equation in the weak form with the test function $\widehat{\phi}_1$ is obtained by calculating the functional derivative
$\int \dd^3\mathbf{x}\ \left(\delta\mathcal{L}/\delta{\phi_1}\right)\cdot\widehat{\phi}_1$:
\begin{eqnarray}
\label{quasin_FLR}
\sum_{\mathrm{s}\neq e} q_s \int \mathrm{d}{\mathbf z}_0\ f_{0,s} (\mathbf{z}_0)\left\langle\widehat{\phi}_1\right\rangle +
e \int  \mathrm{d}{\mathbf z}_0\ f_{0,e} (\mathbf{z}_0)\ \widehat{\phi}_1
\\
\nonumber
\quad\quad =\epsilon_{\delta}\sum_{\mathrm{s}\neq e}\int  \mathrm{d}{\mathbf z}_0\ f_{C,s}(\mathbf{Z}({\bf z}_0,t))\ \frac{m_s c^2}{B^2} \bm\nabla_{\perp}\phi_1\cdot\bm\nabla_{\perp}\widehat{\phi}_1.
\end{eqnarray}
We perform the change of variables $\mathbf{z}=\mathbf{Z}(\mathbf{z}_0,t)$, and the quasineutrality equation becomes
\begin{eqnarray}
\label{quasin_FLR}
\sum_{\mathrm{s}\neq e} q_s \int \mathrm{d}{\mathbf z}\ f_s (\mathbf{z})\left\langle\widehat{\phi}_1\right\rangle +
e \int  \mathrm{d}{\mathbf z}\ f_e(\mathbf{z})\ \widehat{\phi}_1
\\
\nonumber
\quad\quad =\epsilon_{\delta}\sum_{\mathrm{s}\neq e}\int  \mathrm{d}{\mathbf z}\ f_{C,s}(\mathbf{z})\ \frac{m_s c^2}{B^2} \bm\nabla_{\perp}\phi_1\cdot\bm\nabla_{\perp}\widehat{\phi}_1,
\end{eqnarray}
where $\mathrm{d}{\mathbf z}=mB_{\parallel, s} \dd\mathbf{X}\dd p_{z} \dd\mu \dd\theta$.
The Amp\`ere equation obtained from the same variational principle is derived from the computation of $\left(\delta\mathcal{L}/\delta{A_{1\|}}\right)\cdot\widehat{A}_{1\|}$ and the change of variables $\mathbf{z}=\mathbf{Z}(\mathbf{z}_0,t)$:
\begin{eqnarray}
\nonumber
\epsilon_{\delta}\int \frac{\dd^3{\bf x}}{4\pi}\
\bm\nabla_{\perp} A_{1\|}\cdot\bm\nabla_{\perp}\widehat{A}_{1\|}=
\sum_{\mathrm{s}\neq e}\int \ \dd \mathbf{z}\ f_s(\mathbf{z}) \
\frac{q_s\ p_z}{m_s c}\left\langle\widehat{A}_{1\|}\right\rangle
+\int \ \dd \mathbf{z}\ f_e (\mathbf{z}) \
\frac{e\ p_z}{m_e c}\widehat{A}_{1\|}
\\
\nonumber
+\epsilon_{\delta}\sum_{\mathrm{s}\neq e}\int \dd \mathbf{z}\ f_{C,s}(\mathbf{z})\ \left(\frac{q_s^2}{m_s c^2}A_{1\|}\widehat{A}_{1\|}
+\frac{\mu}{B}
\left[
A_{1\|}\bm\nabla_{\perp}^2\widehat{A}_{1\|}+
\widehat{A}_{1\|}\bm\nabla_{\perp}^2A_{1\|}
\right]
\right)
\\
\label{GK_Ampere}
+\epsilon_{\delta}\int \dd \mathbf{z}\ f_{C,e}(\mathbf{z})\ \ \frac{e^2}{m_e c^2}A_{1\|}\widehat{A}_{1\|}.
\end{eqnarray}

%%%%%
\subsection{Nonlinear gyrokinetic Vlasov equation}
%%%%%%

We now proceed with derivation of the particles dynamics implemented in the \orb{} code. 
The equations of motion for the particles are obtained by setting to zero the
functional derivatives with respect to the phase space coordinates
$\mathbf{Z}=\left(\mathbf{X},p_z,\mu,\theta\right)$:
\begin{eqnarray}
\nonumber
\int \dd {\mathbf{z}}_0\  \frac{\delta \mathcal{L}}{\delta \bfZ}\cdot \widehat{ \bfZ}= \sum_{\rm s}\int \dd {\mathbf{z}}_0\ f_{0,s} ({\bf z}_0) \left(   \frac{\partial L_s}{\partial\bfZ} \cdot\widehat{ \bfZ} 
+ \frac{\partial L_s}{\partial\dot{\bfZ}} \cdot \dot{\widehat{ \bfZ} } \right)
 \\
= \sum_{\rm s}\int  \dd \mathbf{z}_0\ f_{0,s}({\bf z}_0)
\left(   \frac{\partial L_s}{\partial \bfZ} \cdot\widehat{ \bfZ}   
-\frac{\dd}{\dt}\frac{\partial L_s}{\partial\dot{\bfZ}}\cdot\widehat{ \bfZ} 
+\frac{\dd}{\dt} \left[  \frac{\partial L_s}{\partial \dot{\bfZ}}\cdot\widehat{ \bfZ} \right] 
\right). 
\label{eq:particles}
 \end{eqnarray}
 Since in the action functional (\ref{L_ORB5}) the nonlinear part of  reduced particle dynamics, i.e., Hamiltonian $H_2$ is only coupled to the non-dynamical part of the distribution function, i.e., $f_C$, $H_2$ does not contribute to the particle characteristics used for reconstructing the gyrokinetic Vlasov equation. 
 The last term in Eq.~(\ref{eq:particles}) vanishes when integrating
 the Lagrangian in time to get the action integral. However, this term is used
 later in order to determine the conserved energy of the system by Noether's theorem. 
 As a consequence, the functional derivatives vanish for all test
 functions $ \widehat{ \bfZ} $ if and only if the Euler-Lagrange
 equation for the particles is satisfied: 
\begin{equation}
\label{eq:E_L}
\frac{ \mathrm{d}}{ \mathrm{d}t}\frac{\partial L_s}{\partial\dot{\bfZ}}=\frac{\partial L_s}{\partial \bfZ}.
\end{equation} 
The gyrokinetic Vlasov equation is reconstructed from the linearised gyrocentre characteristics
according to the approximations performed on the action functional given by Eq.~(\ref{L_ORB5}).
\begin{eqnarray}
0=\frac{\dd f_s}{\dt}=\frac{\partial f_s}{\partial t}
+\dot{\mathbf X}\cdot\bm\nabla f_s+\dot{p_{z}}\frac{\partial f_s}{\partial p_z},
\label{eq:Vlasov_general}
\end{eqnarray}
where the linearised gyrocentre characteristics depend on the linearised Hamiltonian model:
\begin{eqnarray}
\dot{\mathbf X}&=&\frac{c\widehat{\mathbf b}}
{q_s B_{\|,s}^{*}}\times\bm\nabla H_s+\frac{\partial H_s}{\partial{{p}_{z}}}\frac{\mathbf{B}^{*}_s}{B_{\|,s}^{*}},\\
\dot{p}_{z}&=&-\frac{\mathbf{B}^{*}_s}{B_{\|,s}^{*}}\cdot \bm\nabla H_s,
\nonumber
\end{eqnarray}
where $H_s=H_{0,s} +\epsilon_{\delta}H_{1,s}$.
For the ordering considered above, the equations for the ion characteristics become:
\begin{eqnarray}
\label{linear_charakt_ions}
\dot{\mathbf X}&=&\frac{\widehat{\mathbf b}}
{q_s B_{\|,s}^{*}}\times\bm\nabla\left(\mu B+\epsilon_{\delta} q_s \left\langle\phi_1-\frac{p_z}{m_s c}A_{1\|}\right\rangle\right)+\frac{\mathbf{B}_s^{*}}{B_{\|,s}^{*}}\left(\frac{p_z}{m_s}- \epsilon_{\delta}\frac{q_s}{m_s c}\left\langle A_{1\|}\right\rangle\right),\nonumber\\
\dot{p}_{z}&=&-\frac{\mathbf{B}^{*}_s}{B_{\|,s}^{*}}\cdot \bm\nabla \left(\mu B+\epsilon_{\delta} q_s \left\langle\phi_1-\frac{p_z}{m_s c}A_{1\|}\right\rangle\right),
\end{eqnarray}
while the equations for the electron characteristics contain the first FLR corrections only:
\begin{eqnarray}
\label{linear_charakt_el}
\dot{\mathbf X}&=&\frac{\widehat{\mathbf b}}
{e B_{\|,e}^{*}}\times\bm\nabla\left(\mu B+\epsilon_{\delta} e \left( \phi_1-\frac{p_z}{m_e c}A_{1\|}\right)\right)+\frac{\mathbf{B}^{*}_e}{B_{\|,e}^{*}}\left(\frac{p_z}{m_e}- \epsilon_{\delta}\frac{e}{m_e c}\ A_{1\|}\right), \nonumber\\
\dot{p}_{z}&=&-\frac{\mathbf{B}^{*}_e}{B_{\|,e}^{*}}\cdot \bm\nabla \left(\mu B+\epsilon_{\delta} e \left(\phi_1-\frac{p_z}{m_e c}A_{1\|}\right)\right).
\end{eqnarray}
We notice that the equations for the unperturbed characteristics for both species coincide:
\begin{eqnarray}
\label{linear_charakt_nonpert}
\dot{\mathbf X}|_0&=&\frac{\widehat{\mathbf b}}
{q_s B_{\|,s}^{*}}\times\mu\bm\nabla B+\frac{\mathbf{B}^{*}_s}{B_{\|,s}^{*}}\frac{p_z}{m_s}, \nonumber\\
\dot{p}_{z}|_0&=&-\frac{\mathbf{B}^{*}_s}{B_{\|,s}^{*}}\cdot\mu\bm\nabla B.
\end{eqnarray}

%%%%%%%%%%%%%%%%%%%%%%%%%%%%%%%%%%%%%%
\section{\label{sec:Noether_ORB5}Noether theorem for the \orb{} code model}
%%%%%%%%%%%%%%%%%%%%%%%%%%%%%%%%%%%%%%

In order to derive the expression for the energy, we calculate the time derivative of the Lagrangian density $\mathcal {L}=\mathcal{L}(\mathbf{X},p_z,\theta,\mu,\phi,A_{\|})$:
\begin{eqnarray}
\nonumber
\frac{\dd\mathcal L}{\dd t}&=&\int \dd \mathbf{z}_0\left( \frac{\delta\mathcal L}{\delta{\mathbf X}}\cdot\frac{\dd \mathbf X}{\dd t}+\frac{\delta\mathcal L}{\delta{p_z}}  \frac{\dd p_z}{\dd t}+
\frac{\delta\mathcal L}{\delta{ \theta}}  \frac{\dd\theta}{\dd t}+
\frac{\delta\mathcal L}{\delta{\mu}}  \frac{\dd \mu}{\dd t}\right)
\\
&+&\int \dd \mathbf{x}\left(\frac{\delta\mathcal L}{\delta{ \phi_1}} \frac{\partial\phi_1}{\partial t}
+
\frac{\delta\mathcal L}{\delta A_{1\|}} \frac{\partial A_{1\|}}{\partial t}\right).
\end{eqnarray}
Using the Euler-Lagrange equations given by Eq.~(\ref{eq:particles}), we get:
\begin{eqnarray}
\nonumber
\int \dd\mathbf{z}_0\ \frac{\delta\mathcal L}{\delta\mathbf{Z}}\cdot\widehat{\mathbf Z}&=& \int \dd\mathbf{z}_0 f_{0,s}({\bf z}_0)
\left(\frac{\partial L_s}{\partial{\mathbf Z}}\cdot{\widehat{\mathbf Z}}
-\frac{\dd}{\dd t}\ \frac{\partial L_s}{\partial\dot{\mathbf Z}}\cdot\widehat{\mathbf Z}+\frac{\dd}{\dd t}\left[
\frac{\partial L_s}{\partial\dot{\mathbf Z}}\cdot\widehat{\mathbf Z}\right]\right),
\\
&=&
\int \dd\mathbf{z}_0 f_{0,s}({\bf z}_0)\ \frac{\dd}{\dd t}\left(\frac{\partial {L_s}}{\partial\dot{\mathbf Z}}\cdot\widehat{\mathbf{Z}}\right).
\end{eqnarray}
For the field equations, we choose the test function $\widehat{\phi}_1=\partial{\phi}_1/\partial t$ and $\widehat{A}_{1\|}=\partial A_{1\|}/\partial t$ and we use the corresponding Euler-Lagrange equations in the weak form:
\begin{eqnarray}
0=\int \dd\mathbf{x}\ \frac{\delta{\mathcal L}}{\delta\phi_1}\cdot\frac{\partial\phi_1}{\partial t}=\int\dd\mathbf{x}\ \frac{\delta {\mathcal L_s}}{\delta A_{1\|}}\cdot\frac{\partial A_{1\|}}{\partial t}.
\end{eqnarray}
Finally, using the fact that the total time derivative of the Vlasov density vanishes, we get the expression for the energy invariant:
\begin{eqnarray}
\frac{\dd}{\dd t}\left(\mathcal{L}-\sum_{s}\int\dd\mathbf{z}_0\  f_{0,s}({\mathbf{z}}_0)\left[ \frac{\partial{L_s}}{\partial\dot{\mathbf X}}\cdot{\mathbf{\widehat{X}}}-\frac{\partial {L_s}}{\partial\dot{\theta}}\cdot{\widehat{\theta}}\   \right] \right)=\frac{\dd}{\dd t}{\mathcal E}^{\mathrm{EM}}=0, 
\end{eqnarray}
where
\begin{eqnarray*}
\label{E_em}
\mathcal{E}^{\mathrm{EM}}= \sum_{\mathrm{s}} \int \dd\mathbf{z}_0\  f_{0,s}(\mathbf{z}_0) \left(H_{0,s}+\epsilon_{\delta} H_{1,s}\right) &+& \epsilon_{\delta}^2 \sum_{\mathrm{s}} \int \dd\mathbf{z}_0 \ f_{C,s}(\mathbf{Z}(\mathbf{z}_0;t)) \  H_{2,s}
\nonumber
\\
&+&
\epsilon_{\delta}^2\int \dd{\bf x} \frac{\left|\bm\nabla_{\perp} A_{1\|}\right|^2}{8\pi}.
\end{eqnarray*}
Using the same change of variables ${\mathbf z}=\mathbf{Z}(\mathbf{z}_0;t)$, as in the Poisson and Amp\`ere equations, we get: 
\begin{eqnarray}
\label{E_em}
\mathcal{E}^{\mathrm{EM}}= \sum_{\mathrm{s}} \int \dd\mathbf{z}\  f_s(\mathbf{z}) \left(H_{0,s}+\epsilon_{\delta} H_{1,s}\right) &+& \epsilon_{\delta}^2 \sum_{\mathrm{s}} \int \dd\mathbf{z} \ f_{C,s}(\mathbf{z}) \  H_{2,s}
\nonumber
\\
&+&
\epsilon_{\delta}^2\int \dd{\bf x} \frac{\left|\bm\nabla_{\perp} A_{1\|}\right|^2}{8\pi}.
\end{eqnarray}
The procedure allowing one to get the power balance diagnostics is the following one: First, we directly substitute the expression for Hamiltonians $H_{0,s}$, $H_{1,s}$ given by Eqs.~(\ref{H_{0,s}}-\ref{H_1}) and $H_2$ given by Eqs.~(\ref{H_2FLR}-\ref{H_2_e}). Then we define the unperturbed kinetic energy of the particles:
$${\mathcal E}_{\mathrm{kin}}=\sum_{\mathrm{s}} \int \dd\mathbf{z} \ f_s(\mathbf{z})\ H_{0,s} =\sum_{\mathrm{s}}\int \dd\mathbf{z}\ f_s(\mathbf{z}) \left(\frac{p_z^2}{2 m_s}+\mu B\right), 
$$ 
and the remaining terms are referred to as field energy:
\begin{eqnarray}
\label{E_F_raw}
{\mathcal E}_F&=& \epsilon_{\delta}\sum_{\mathrm{s}} \int \dd\mathbf{z}\ f_s(\mathbf{z}) H_{1,s} +\epsilon_{\delta}^2 \sum_{\mathrm{s}} \int\dd\mathbf{z}\ f_{C,s}(\mathbf{z}) H_{2,s} + \ 
\epsilon_{\delta}^2\int{\mathrm d}\mathbf{x}\frac{\left|\bm\nabla_{\perp} A_{1\|}\right|^2}{8\pi},
\\
\nonumber
&=&\epsilon_{\delta} \sum_{\mathrm{s}\neq e}\int \dd\mathbf{z}\ f_s(\mathbf{z})\ q_s\left\langle\phi_1- A_{1\|}\frac{p_{z}}{m_s c} \right\rangle +
\epsilon_{\delta} \int \dd\mathbf{z}\ f_e(\mathbf{z})\ 
e \left( \phi_1- A_{1\|}\frac{p_{z}}{m_e c}\right) 
\\
\nonumber
&+&\epsilon_{\delta}^2\sum_{\mathrm{s}\neq e}\int \dd\mathbf{z}\ f_{C,s}(\mathbf{z}) \left[-\frac{m_s c^2}{2B^2}\left|\bm\nabla_{\perp}\phi_1\right|^2+\frac{q_s^2 }{2 m_s c^2}\left(A_{1{\|}}^2+\left(\frac{m_s c^2}{q_s^2}\right) \frac{\mu}{B}A_{1\|}\bm\nabla_{\perp} ^2A_{1\|}\right)\right] 
\\
\nonumber
&+&\epsilon_{\delta}^2\int \dd\mathbf{z} \ f_{C,e}(\mathbf{z})\ \frac{e^2 }{2 m_e c^2} A_{1{\|}}^2+
\epsilon_{\delta}^2\int \mathrm{d}{\mathbf x}\ \frac{\left|\bm\nabla_{\perp} A_{1\|}\right|^2}{8\pi}.
\end{eqnarray}
 Next, the nonlinear term containing $H_2$ in the expression for the energy is rewritten using the corresponding quasineutrality and Amp\`ere equations in the weak form. This is achieved by choosing a particular test function $\widehat{\phi}_1=\phi_1$ and by substituting it in Eq.~(\ref{quasin_FLR}). Similarly, the test function $\widehat{A}_{1\|}=A_{1\|}$ is substituted to the corresponding Amp\`ere equation given by Eq.~(\ref{GK_Ampere}).
 The quasineutrality equation (\ref{quasin_FLR}) with $\widehat{\phi}_1=\phi_1$ is written as:
\begin{eqnarray}
\label{weak_quasin_FLR}
&&\epsilon_{\delta} \sum_{s}\int \dd\mathbf{z}\ f_{C,s}(\mathbf{z})\ \frac{m_s c^2}{B^2} \left|\bm\nabla_{\perp}\phi_1\right|^2
=
\\
\nonumber
&&\quad\quad\quad \sum_{\mathrm{s}\neq e}\int \dd\mathbf{z}\ f_s(\mathbf{z})\ q_s\left\langle{\phi}_1\right\rangle+
\int \dd\mathbf{z}\ f_e(\mathbf{z})\ e\ {\phi}_1.
\end{eqnarray}
The Amp\`ere  equation~(\ref{GK_Ampere}) with  $\widehat{A}_{1\|}=A_{1\|}$ is written as:
\begin{eqnarray}
\nonumber
\epsilon_{\delta}\int \frac{\mathrm{d}\mathbf{x}}{4\pi}\ 
\left|\bm\nabla_{\perp} A_{1\|}\right|^2
&=&
\sum_{\mathrm{s}\neq e}\int \dd\mathbf{z}\ f_s(\mathbf{z}) \ 
\frac{p_z}{m_s}\left\langle A_{1\|}\right\rangle+
\int \dd\mathbf{z}\ f_e(\mathbf{z}) \ 
\frac{p_z}{m_e}\ A_{1\|}
\\
\nonumber
&-&\sum_{\mathrm{s}\neq e}\epsilon_{\delta}\int \dd{\mathbf z}\ f_{C,s}(\mathbf{z})\ \left(\frac{q_s^2}{m_s}A_{1\|}^2
+\frac{\mu}{B}
A_{1\|}\bm\nabla_{\perp}^2 A_{1\|}
\right)
\\
&-&
\label{weak_Ampere_GK}
\epsilon_{\delta}\int\dd\mathbf{z}\ f_{C,e}(\mathbf{z})\ \frac{e^2}{m_e}A_{1\|}^2.
\end{eqnarray}
Now using Eqs.~(\ref{weak_quasin_FLR}-\ref{weak_Ampere_GK})  we substitute the expressions for the electrostatic and electromagnetic contributions into Eq.~(\ref{E_F_raw}) and we get Eq.~(\ref{E_EM_FULL}):
\begin{eqnarray}
\mathcal{E}^{\mathrm{EM}}&=&\frac{1}{2}\sum_{\mathrm{s}\neq e}q_s\int \dd\mathbf{z}\ f_s(\mathbf{z}) \left\langle\phi_1-\frac{p_z}{m_s c} A_{1\|}\right\rangle  +
\frac{e}{2}\int \dd\mathbf{z}\ f_e(\mathbf{z})\ \left(\phi_1-\frac{p_z}{m_e c}\ A_{1\|} \right) 
\nonumber
\\
&+&
\sum_{\mathrm{s}}\int \dd\mathbf{z}\ f_s(\mathbf{z})\left(\frac{p_z^2}{2 m_s}+\mu B\right)\equiv\mathcal{E}_F+\mathcal{E}_{\mathrm{kin}}.
\label{E_EM_FULL}
\end{eqnarray}
For clarity, we define a function for each component of $\mathcal{E}_{\mathrm{EM}}$:
\begin{eqnarray}
\label{E_F_decomposition}
\mathcal{E}_F\equiv\mathcal{E}_{\mathrm{es}}-\mathcal{E}_{\mathrm{em}},\\
\label{E_es}
\mathcal{E}_{\mathrm{es}}=\frac{1}{2}\sum_{\mathrm{s}\neq e}q_s\int \dd\mathbf{z}\ f_s(\mathbf{z}) \left\langle\phi_1\right\rangle\ +\frac{e}{2}\int \dd\mathbf{z}\ f_e(\mathbf{z})\ \phi_1,\\
\label{E_em}
\mathcal{E}_{\mathrm{em}}=\frac{1}{2}\sum_{\mathrm{s}\neq e}q_s\int\dd\mathbf{z}\ f_s(\mathbf{z}) \ \frac{p_z}{m_s c}\left\langle A_{1\|}\right\rangle\ 
+\frac{e}{2}\int \dd\mathbf{z}\ f_e(\mathbf{z}) \frac{p_z}{m_e c} A_{1\|},\\
\label{E_kin}
\mathcal{E}_{\mathrm{kin}}=\sum_{\mathrm{s}}\epsilon_{\delta}\int \dd\mathbf{z}\ f_s(\mathbf{z})\ \left(\frac{p_z^2}{2 m_s}+\mu B\right).
\end{eqnarray}
We remark that these expressions are general for all electromagnetic models and is independent from the choice of the nonlinear model, i.e., the second order Hamiltonian $H_2$.

%%%%
\section{\label{sec:PB_diag}Energetically consistent diagnostics for the \orb{} code }
%%%%

The derivation of dynamical invariants via the Noether's method is naturally included in the Lagrangian framework. It gives an opportunity to construct code diagnostics, allowing one to control the quality of the simulations and to get information about  the mechanisms triggering the instabilities.

In particle-in-cell codes, the dynamics of particles and fields is computed in two different ways: Particles are advanced along their characteristics without the use of any grid, while fields are evaluated on a grid. Within one calculation cycle, both sides are communicating: Particle are pushed along their renewed characteristics by using the values of the electromagnetic fields evaluated on the grid. Then the new values of the particle positions are deposed on the grid in order to provide the new values for the charge and current density entering into the electromagnetic field equations (\ref{quasin_FLR})-(\ref{GK_Ampere}).

Considering the energy exchange between particles and fields, i.e., independently calculating the time derivatives of $\mathcal{E}_{\mathrm{kin}}$ and $\mathcal{E}_F$, allows one to control the consistency of the algorithm and the quality of the simulation by verifying the energy conservation. Moreover, further application of Noether theorem makes it possible to analytically calculate a simplified expression for the time derivative of the field energy $\mathcal{E}_F$,  given by the non-perturbed characteristics of the particles only. Such a simplification gives a possibility to access  the underlying instability mechanisms through the particle characteristics.

 In this section, we provide the detailed derivation of the \orb{} diagnostics developed from the field-particles energy balance equation:
\begin{equation}
\frac{{\mathrm d}{\mathcal E}_{\mathrm{kin}}}{{\mathrm d} t}=-\frac{{\mathrm d}{\mathcal E}_{\mathrm{F}}}{{\mathrm d} t},
\label{dE_k_dE_F_1}
\end{equation}
where the time derivative of the l.h.s.\ can be evaluated through the particle characteristics and the r.h.s.\ from the field contributions evaluated on the grid.

Two diagnostics issued from  Eq.~(\ref{dE_k_dE_F_1}) are defined: First, the power balance diagnostics is defined as the energy balance equation (\ref{dE_k_dE_F_1}), normalised by the field energy $\mathcal{E}_F$. Second, the $\Delta \mathcal{E}_F$ diagnostics is defined as the energy balance equation (\ref{dE_k_dE_F_1}) normalised by an electrostatic component of the field energy, i.e., $\mathcal{E}_{\mathrm{es}}$. 

Using the definition of the kinetic part of the energy $\mathcal{E}_{\mathrm{kin}}$, given by Eq.~(\ref{E_kin}), we explicitly calculate the contributions to Eq.~(\ref{dE_k_dE_F_1}):
\begin{eqnarray*}
\frac{\dd \mathcal{E}_{\mathrm{kin}}}{\dt}&=&-\frac{1}{2}\sum_{{\mathrm s}\neq e}\ q_s\int\dd\mathbf{z}\ f_s(\mathbf{z})\left(\dot{\mathbf{X}}\cdot\bm\nabla\left\langle\psi_{1,s}\right\rangle-
\frac{\dot{p_z}}{m_s c}\left\langle A_{1\|}\right\rangle\right)
\\
&-&\frac{e}{2}\int\dd\mathbf{z}\ f_e(\mathbf{z})\left(\dot{\mathbf{X}}\cdot\bm\nabla\psi_{1,e}-
\frac{\dot{p_z}}{m_e c} A_{1\|}\right) 
=-\frac{\dd \mathcal{E}_F}{\dt},
\end{eqnarray*}
where
\begin{equation}
\psi_{1,s}=\phi_1-A_{1\|}\frac{p_{z}}{m_s c}.
\end{equation}
The direct implementation of the linear gyrocentre characteristics for $\mathbf X$ and $p_z$ given by 
Eqs.~(\ref{linear_charakt_ions}-\ref{linear_charakt_el}) leads to the cancellation of all nonlinear terms related to the perturbed electromagnetic fields, i.e., the final expression contains only the contributions corresponding to the unperturbed Hamiltonian dynamics given by Eq.~(\ref{linear_charakt_nonpert}):
\begin{eqnarray}
\frac{\dd \mathcal{E}_{\mathrm{kin}}}{\dt}&=&-\frac{\dd \mathcal{E}_F}{\dt}=-\frac{1}{2}\sum_{\mathrm s\neq e} q_s\int\dd\mathbf{z}\ f_s(\mathbf{z})\left(\dot{\mathbf{X}}|_{0}\cdot\bm\nabla\left\langle\psi_{1,s}\right\rangle-
\frac{\dot{p_z}|_0}{m_s c}\left\langle A_{1\|}\right\rangle\right)
\nonumber
\\
&-&
\frac{e}{2} \int\dd\mathbf{z}\ f_e(\mathbf{z})\left(\dot{\mathbf{X}}|_{0}\cdot\bm\nabla\psi_{1,e}-
\frac{\dot{p_z}|_0}{m_e c}\ A_{1\|}\right),
\label{energy_balance_kin}
\end{eqnarray}
where
\begin{eqnarray}
\label{unperturbed_charakt}
\dot{\mathbf X}\left|_0\right.&=&\frac{p_z}{m_s} \widehat{\mathbf b}+\frac{c\widehat{\mathbf b}}
{q_s B_{\|,s}^{*}}\times\left(\mu\bm\nabla B-\left(\frac{p_z^2}{m_s}\ \widehat{\mathbf b}\times\bm\nabla\times{\widehat{\mathbf b}}\right)\right).
\end{eqnarray}
The geometric contribution $\bm\nabla\times\widehat{\mathbf  b}$  to ${\mathbf B}_s^*$ given by Eqs.~(\ref{eq:B*},~\ref{eq:A*}) is expressed by using the projection on the parallel and perpendicular directions, following the calculations given in Appendix B of \cite{TBS_2016}:
\begin{eqnarray*}
\bm\nabla\times\widehat{\mathbf b}=\widehat{\mathbf b}\left(\widehat{\mathbf b}\cdot\bm\nabla\times\widehat{\mathbf b}\right)-
\widehat{\mathbf b}\times\left[\widehat{\mathbf b}\times\bm\nabla\times\widehat{\mathbf b}\right]\equiv \widehat{\mathbf b}\ \tau- {\mathbf G},
\end{eqnarray*}
where the scalar $\tau$ represents the magnetic twist and the vector $\mathbf G$ is referred to as the magnetic curvature.
Since $\mathbf B\times \left(\bm\nabla\times{\mathbf B}\right)= -\bm\nabla p$ in single fluid MHD equilibrium, 
we rewrite the curvature vector $\mathbf G$ in order to evidence the pressure-like contributions into the characteristics:
\begin{eqnarray*}
\mathbf G=\widehat{\mathbf b}\times\left[\widehat{\mathbf b}\times\frac{\bm\nabla\times{\mathbf B}}{B}\right] +
\frac{\bm\nabla B\times\widehat{\mathbf b}}{B}=-\widehat{\mathbf b}\times\left(\frac{\bm\nabla p}{B^2}+\frac{\bm\nabla B}{B}\right).
\end{eqnarray*}
We also decompose the geometric magnetic field in the parallel and perpendicular components in the following way:
\begin{eqnarray*}
\mathbf B_s^*= \underbrace{\left(B +p_z\frac{c}{q_s} \ \tau \right)}_{\equiv B_{\|,s}^*}\widehat{\mathbf b} -p_z\frac{c}{q_s} \mathbf{G}.
\end{eqnarray*}
The unperturbed characteristics in the power balance equation are given by:
\begin{eqnarray}
\dot{\mathbf X}|_0&=&\frac{p_z}{m_s}\ \widehat{\mathbf b}+\left(\frac{p_z}{m_s}\right)^2\frac{m_s}{q_s B_{\|,s}^*}\widehat{\mathbf b}
\times\left(\frac{\bm\nabla p}{B^2}\right)
+\left(\frac{\mu}{m_s}+\left(\frac{p_z}{m_s}\right)^2\right)\frac{m_s}{q_s B_{\|,s}^*}
\widehat{\mathbf b}\times\frac{\bm\nabla B}{B},
\nonumber
\\\dot{p}_z|_0&=&\mu B\ \bm\nabla\cdot\widehat{\mathbf b} + \frac{\mu c}{q_s B_{\|,s}^*}\ p_z\left( \widehat{\mathbf b}\times
\frac{\bm\nabla B}{B^2}\right)\cdot\bm\nabla p,
\label{GC_pz}
\end{eqnarray}
where we have used the divergence free property of magnetic field: $\widehat{\mathbf b}\cdot\bm\nabla B=- B\ \bm\nabla\cdot\widehat{\mathbf b}$.

%%%%%%%%
\subsection{\label{sec:code_implmentation}The power balance diagnostics}
%%%%%%%

In order to understand and analyse possible sources of plasma deconfinement, one aims to investigate mechanisms, triggering the growth of microinstabilities and turbulent transport.
The mechanisms contributing to the development of microinstabilities are directly related to the exponential growth of electromagnetic field fluctuations.
Considering that  electromagnetic instabilities have an exponential growth: $\mathcal{E}_F=\bar{\mathcal{E}}_F e^{\gamma t}$, we derive the expression for the power balance diagnostics:
\begin{equation}
\frac{\dd \mathcal{E}_F}{\dt}=\gamma\mathcal{E}_F\; \Rightarrow\; \gamma=\frac{1}{\mathcal{E}_F}\frac{\dd \mathcal{E}_F}{\dt}.
\label{gamma}
\end{equation}
Therefore, for practical reasons, in numerical simulations, it is useful to consider the power balance equation in the following form (i.e., normalized by the field energy $\mathcal E_F$):
\begin{eqnarray}
\frac{1}{\mathcal E_F}\frac{\dd\mathcal E_{\rm kin}}{\dd t}=-\frac{1}{\mathcal E_F}\frac{\dd\mathcal E_F}{\dd t}.
\label{energy_balance_norm}
\end{eqnarray}
The power balance diagnostics is suitable for quality verification in linear and nonlinear simulations.
In addition to that in the case of the linear simulations, the power balance equation not only gives an indication about the quality of the simulation but also can be used to measure the instantaneous growth rate of instability \cite{TBS_2016}.

%%%%%%%
\subsection{$\Delta{\mathcal E}_F$ diagnostics}
%%%%%%%

In the case of electrostatic simulations, the power balance diagnostics is sufficient for investigating the stabilizing and destabilizing mechanisms. The situation is slightly different in the case of electromagnetic simulations when $\mathcal{E}_{\mathrm{es}}=\mathcal{E}_{\mathrm{em}}$ therefore, $\mathcal{E}_F=0$ and the diagnostics defined by Eq.~(\ref{energy_balance_norm}) is not defined.
In order to investigate the transition between electrostatic and electromagnetic instabilities, we introduce the following diagnostics:
\begin{equation}
\label{Delta_E_F_1}
\Delta\mathcal{E}_F=\frac{1}{\mathcal{E}_{\mathrm{es}}}\frac{\dd{\mathcal E}_{\mathrm{es}}}{\dd t}-\frac{1}{\mathcal{E}_{\mathrm{es}}}\frac{\dd{\mathcal E}_{\mathrm{em}}}{\dd t},
\end{equation}
where functions ${\mathcal E}_{\mathrm{es}}$ and ${\mathcal E}_{\mathrm{em}}$  are defined accordingly to Eqs.~(\ref{E_es})-(\ref{E_em}).

This diagnostics allows one to investigate the properties of the electromagnetic simulations from different viewpoints:
First, the sign of the function $\Delta{\mathcal{E}_F}$ determines if the instability is electrostatic (positive) or electromagnetic (negative). Moreover, $\Delta{\mathcal{E}}_F$ allows one to access all stabilizing/destabilizing mechanisms through Eq.~(\ref{energy_balance_kin}) even in the situation with $\mathcal{E}_F=0$.
In addition, investigating functional properties of $\Delta{\mathcal{E}}_F$ as a function of magnetic $\beta$, i.e., the zeros and extremum points (minimum for example) gives the possibility to analyse bifurcations and exchanges of stabilizing/destabilizing mechanisms.
We provide this analysis for linear electromagnetic simulations in the next section.

%%%%%%%%
\section{\label{sec:num_analysis}Analysis of electromagnetic simulations }
%%%%%%%
\subsection{Numerical setup}
%%%%%%%

The parameters used in simulations are derived from the Cyclone Base Case (CBC), which is a well established set of parameters for the flux tube (simulations for one magnetic field line) and global studies (simulations covering the full radial range in the small section of the device). First, it has been used \cite{Dimits_2000} for the benchmark of different flux tubes codes and recently for the benchmark of global electromagnetic codes \cite{Goerler_Tronko_2016}. The original discharge (H- mode shot $\#81499$ taken at $t=4000$ ms and minor radius $r=0.5a$, where $a$ is the the minor radius) of the DIII-D device which serves as a basis for the CBC has naturally more complex shaped flux surfaces. In our case, the equilibrium magnetic configuration is circular and concentric with the inverse aspect ratio $a/R_0=0.36$ and the safety factor profile:
\begin{equation}
q(r)=2.52\left(r/a\right)^2-0.16\left(r/a\right)+0.86.
\end{equation}
Here $a$ is the minor tokamak radius and $R_0$ is a major one, $r$ is the local radius of a flux surface.
The temperature and density profiles and their normalized logarithmic gradients are given by:
\begin{eqnarray}
A/A(r_0)&=&\exp\left[-\kappa_A w_A \frac{a}{L_{\mathrm{ref}}}\tanh\left(\frac{r-r_0}{w_A a}\right)\right],\\
L_{\mathrm{ref}}/L_A&=&-L_{\mathrm{ref}}\partial_r \ln A(r)=\kappa_A\cosh^{-2}\left(\frac{r-r_0}{w_A a}\right),
\end{eqnarray}
which gives us a peaked gradient profile of density and temperature $A=(n,T)$ centred at $r=r_0$ with maximal amplitude $\kappa_A$ and characteristic width $w_A$. The macroscopic reference length $L_{\mathrm{ref}}$ is fixed to the major radius $R_0$ in what follows.

The values of parameters used for the benchmark are summarized in Table \ref{param_1}. We remark that the profiles for ions and electrons are chosen to be identical.

\begin{table}[h]
\begin{center}
\begin{tabular}{|l|c|}
\hline
$r_0/a$&$0.5$\\
$a/L_{\mathrm{ref}}$&$0.36$\\
$R_0/L_{\mathrm{ref}}$&$1.0$\\
\hline
$T_i(r_0)/T_{\mathrm{ref}}=T_e(r_0)/T_{\mathrm{ref}}$&$1.0$\\
$\kappa_{T_i}=\kappa_{T_e}$&$6.96$\\
$w_{T_i}=w_{T_e}$&$0.3$\\
\hline
$n_i(r_0)/T_{\mathrm{ref}}=n_e(r_0)/T_{\mathrm{ref}}$&$1.0$\\
$\kappa_{n_i}=\kappa_{n_e}$&$2.23$\\
$w_{n_i}=w_{n_e}$&$0.3$\\
\hline
$m_i/m_{\mathrm{ref}}$&$1.0$\\
$m_e/m_{\mathrm{ref}}$&$5.44617\cdot 10^{-4}$\\
\hline
\end{tabular}
\end{center}
\caption{\label{param_1} List of benchmark parameters}
\end{table}
The nominal reference values issued from the original experimental work \cite{Greenfield_1997} are given in Table ~\ref{max_amplitudes}.
\begin{table}[h!]
\begin{center}
\begin{tabular}{|l|c|}
\hline
$m_{\mathrm{ref}} (=m_D)/m_p$&$2.0$\\
$n_{\mathrm{ref}}(=n_e)/10^{19} m^{-3}$&$4.66$\\
$T_{\mathrm{ref}}(=T_e)/\mathrm{KeV}$&$2.14$\\
$B_{\mathrm{ref}}(=B_t(R_{\mathrm{mag}}))/T$&$2.0$\\
$L_{\mathrm{ref}}(=R_0=R_{\mathrm{mag}})/m$&$1.67$\\
\hline
$\beta_{\mathrm{ref}}$&$0.0045$\\%$0.0101$\\
$\rho^*=\rho_s/a$&$0.00555\sim1/180.2$\\
\hline
\end{tabular}
\end{center}
\caption{\label{max_amplitudes} Nominal reference and derived reference values based on the low elongation magnetic surfaces case (CBC) \cite{Greenfield_1997}, Fig.5, discharge $\#81499$, at time $t=4000$ms and $\rho=0.5$, which after the rescaling of magnetic surfaces shape towards concentric surfaces corresponds to
$r/a=0.5$.}
\end{table}

In order to reduce the resolution requirement and the computational effort, the ion-electron mass ratio is set to the proton-electron mass ratio, i.e. the electrons are considered being twice heavier than in reality. Concerning the spatial resolution, the associated finite-size parameter $\rho^*=\rho_i/a$, defined as the ratio between the ion gyroradius $\rho_i$ and the minor radius $a$, is set to $1/180$. In fact, considering a hydrogen aspect ratio for ions would need smaller spatial scales by a factor 2. 

The nominal $\beta$ value at the reference position $r/a=0.5$ is close to $0.5\%$ assuming we have taken in consideration the following definition of
$\beta_{\mathrm{ref}}=8\pi n_{\mathrm{ref}}  T_{\mathrm{ref}}/(B^2_{\mathrm{ref}})$ and the normalisation of $n_{\rm{ref}}$, corresponding to the \orb{} code. Here $n_{\mathrm{ref}}$ and $T_{\mathrm{ref}}$ are the density and temperature taken at the reference position  $r/a$.

%%%%%%%%%%%%
\subsection{Electrostatic and electromagnetic instability analysis}
%%%%%%%%%%%%

Introducing the electromagnetic effects into the gyrokinetic simulations adds significant complexity compared to the electrostatic simulations, and requires a more detailed analysis for the implementation of diagnostics. It has not been remarked in electrostatic simulations that the normalization of the power balance diagnostics (\ref{gamma}) on the field energy $\mathcal{E}_F$ together with using this diagnostics for the growth rate calculation in linear simulations may introduce some inconsistencies. It starts to be evident for the electromagnetic simulations when the amount of magnetic energy is equal to the amount of electrostatic energy, i.e., $\mathcal{E}_F=0$. It is evident that in this case, Eq.~(\ref{gamma}) cannot be used for measuring the growth rate. However, thanks to a small modification, this diagnostic can be adapted for investigating the behaviour of the instability triggering mechanisms at this transitional point [see Eq.~(\ref{Delta_E_F_1})].

The contributions to the growth rate $\gamma$ arising from the different terms in the unperturbed guiding-center characteristics 
$\dot{\mathbf X}|_0$ and $\dot{p}_z |_0$ can be separated in the power balance equation and give a clear vision of which type of instability is present in the system: This diagnostics is suitable for both linear and nonlinear electromangetic simulations:
\begin{eqnarray}
\frac{\dd{\mathcal E}_F}{\dt}
&=& 
\frac{1}{2}\sum_{\mathrm{s}\neq e}\int \dd\mathbf{z}\ f_s(\mathbf{z}) \bm\nabla \left\langle\psi_{1,s}\right\rangle\cdot
\left(\bm{v}_{\|,s}+\bm{v}_{{{\bm\nabla} P},s}+\bm{v}_{{{\bm\nabla} B},s}\right)
\nonumber
\\
&+&\frac{1}{2}\int \dd\mathbf{z}\ f_e(\mathbf{z}) \bm\nabla \psi_{1,e} \cdot
\left(\bm{v}_{\|,e}+\bm{v}_{{{\bm\nabla} P},e}+\bm{v}_{{{\bm\nabla} B},e}\right)
\\
\nonumber
&-&\frac{1}{2}\sum_{\mathrm{s}\neq e}\int \dd\mathbf{z}\ f_s(\mathbf{z}) \left\langle A_{1\|}\right\rangle \bm{v}_{\mathrm{curv},s}
-\frac{1}{2}\int \dd\mathbf{z}\ f_e(\mathbf{z}) \ A_{1\|}\ \bm{v}_{\mathrm{curv},e},
\label{dEf_dt}
\end{eqnarray}
where
\begin{eqnarray}
\bm{v}_{\|,s}\equiv\frac{p_z}{m_s}\widehat{\mathbf b}, \\
\bm{v}_{\bm{\nabla}P,s}\equiv-\left(\frac{p_z}{m_s}\right)^2\frac{m_s c}{q_s B_{\|,s}^*}\widehat{\mathbf b}\times\frac{\bm\nabla P}{B^2}, \\
\bm{v}_{\bm{\nabla}B,s}\equiv \left(\frac{\mu B}{m_s}+
\left(\frac{p_z}{m_s}\right)^2\right)\frac{m_s}{q_s B_{\|,s}^*}\widehat{\mathbf b}\times\frac{\bm\nabla B}{B}, \\
\bm{v}_{\mathrm{curv},s}=
\mu B \bm\nabla\cdot\widehat{\mathbf b} + \frac{\mu c}{q_s B_{\|,s}^*}\ p_z \left(\widehat{\mathbf b}\times\frac{\bm\nabla B}{B^2}\right)\cdot\bm\nabla P.
\end{eqnarray}
Consistently with the choice of magnetic equilibrium effects, we are neglecting the $\bm\nabla P$ contribution to the curvature or $\bm\nabla B$ drift. Therefore,  $\bm{v}_{{\bm\nabla P},s}=0$ and $\bm{v}_{\mathrm{curv},s}=
\mu B \bm\nabla\cdot\widehat{\mathbf b}$. This gives us the possibility to follow easily the dynamics of the remaining contributions to $\dd\mathcal{E}_F/ \dt$.
Accordingly to the sign, the contributions $\bm{v}_{\|,s}$, $\bm{v}_{\rm{curv},s}$ and $\bm{v}_{{\bm{\nabla}B},s}$ to the time derivative of the field energy are considered  stabilizing when it is negative or destabilizing when it is positive.

Following the electromagnetic $\beta$ scan of the linear electromagnetic simulations, summarized in the Table~\ref{tab3}, we focus on the sign of $\bm{v}_{\|,s}$,$\bm{v}_{\rm{curv},s}$ and $\bm{v}_{\bm\nabla B,s}$ contributions. In Fig.~1, different cases of the instability triggering mechanisms are presented. 
We notice that for the case with $\mathcal{E}_F=0$ or close to zero, additional standard linear fit diagnostics for the growth rate is used to avoid numerical errors due to the division by a small number in the denominator of Eq.~(\ref{gamma}).
At the same time, for each value of $\beta$, we monitor the value of $\Delta E_F$, defined by Eq.~(\ref{Delta_E_F_1}), as the value of the electromagnetic field energy normalized by the electrostatic energy (see Fig.~2). The change of sign for  $\Delta E_F$ corresponds to the transition from the electrostatic to the electromagnetic regime.
\begin{figure}[h!]
\begin{center}
\begin{tabular}{c c}
\includegraphics[height=4cm,width=7cm]{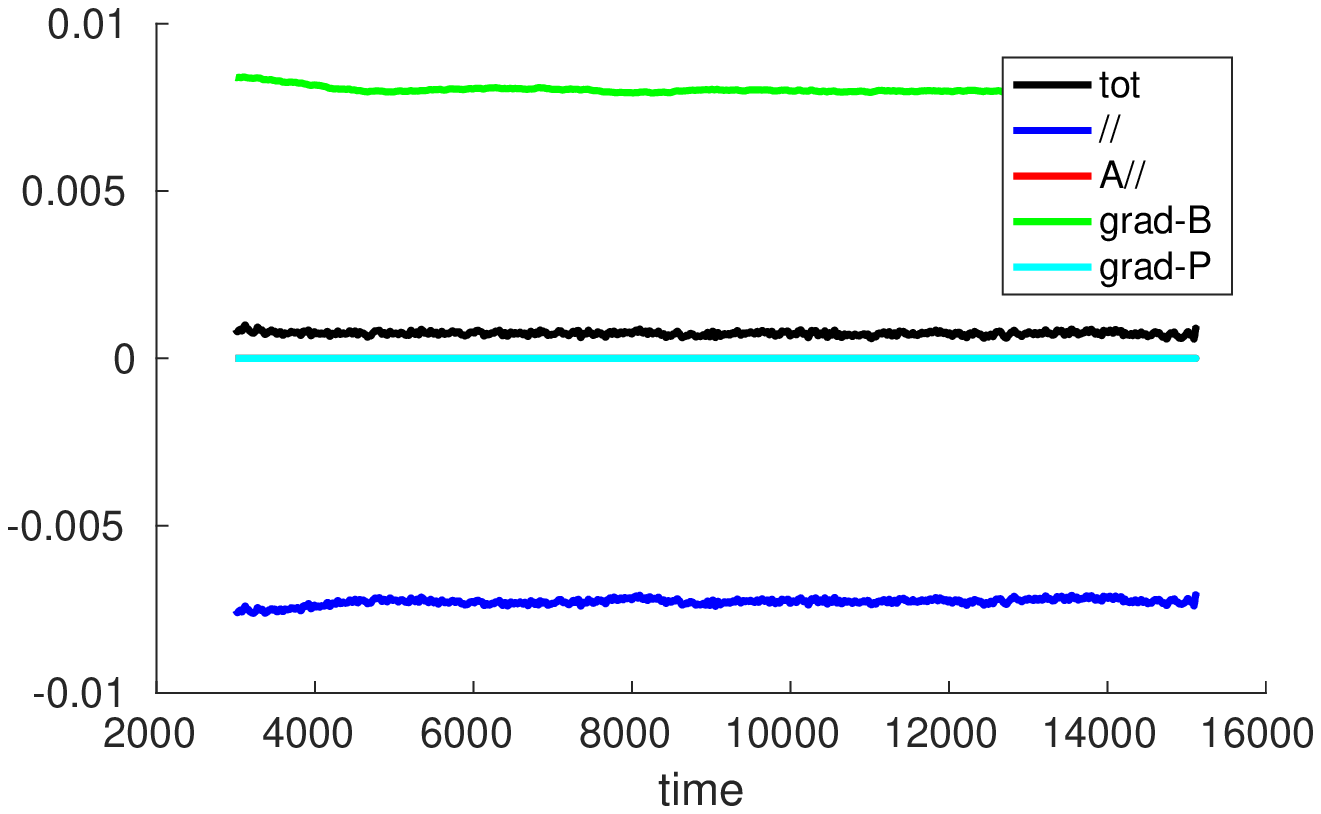}
\label{ITG_small}
&
\includegraphics[height=4cm,width=7cm]{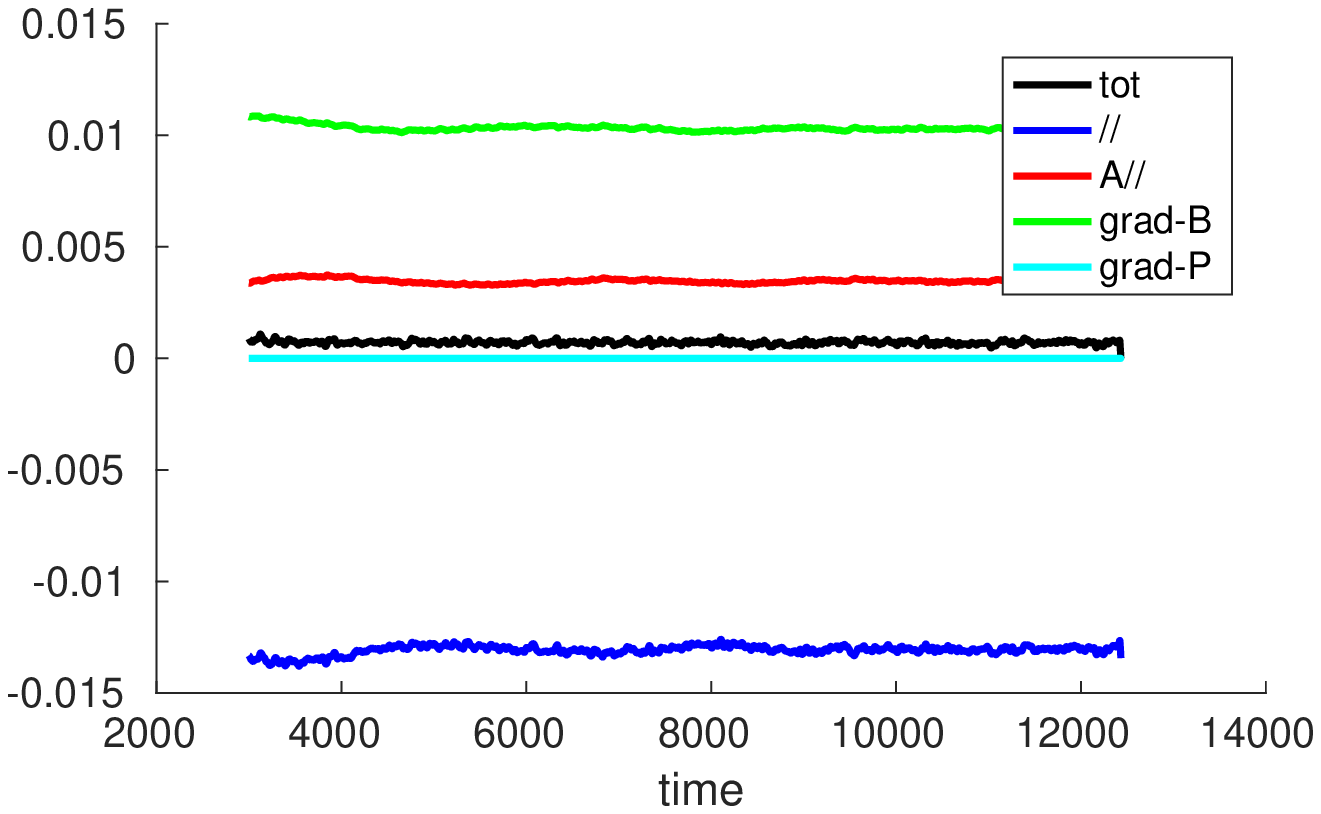}
\label{ITG_00025}
\\
a)&b)
\\
&
\\
\includegraphics[height=4cm,width=7cm]{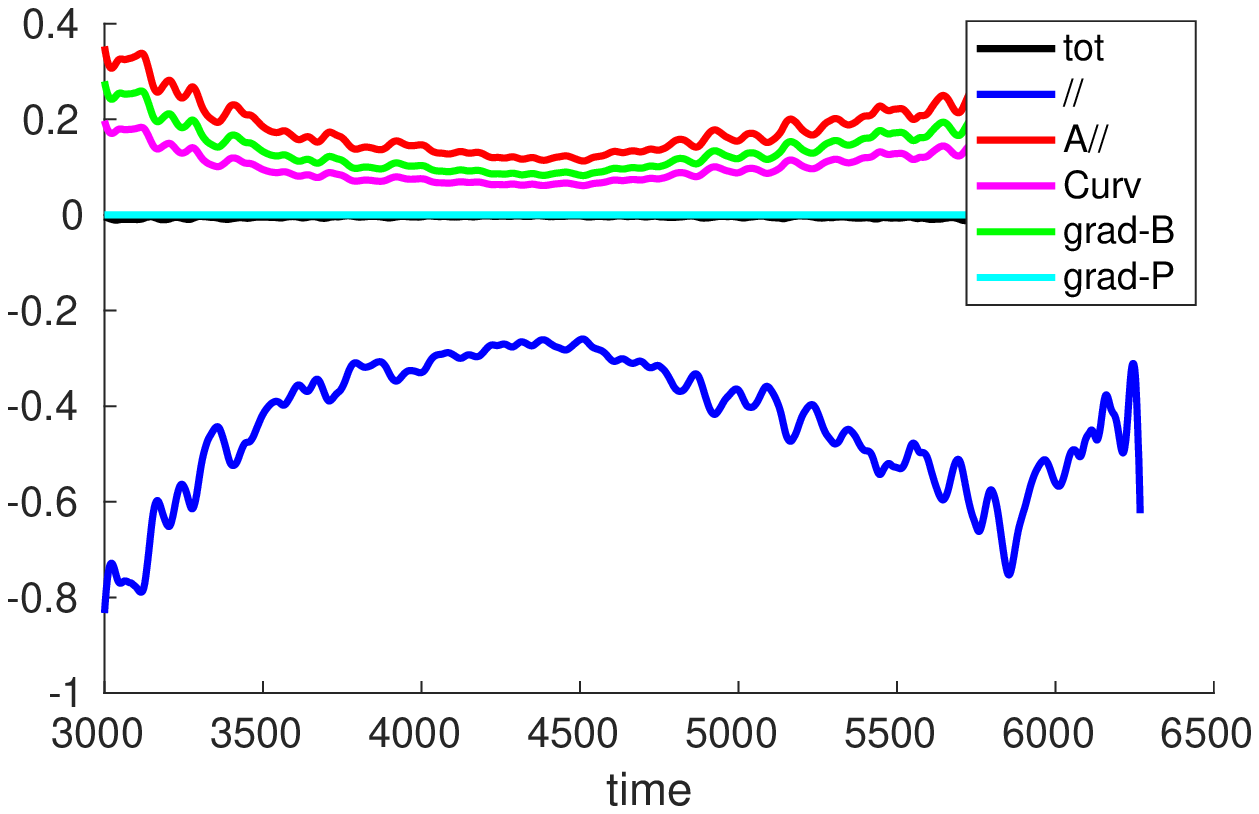}
\label{E_F=0}&
\includegraphics[height=4cm,width=7cm]{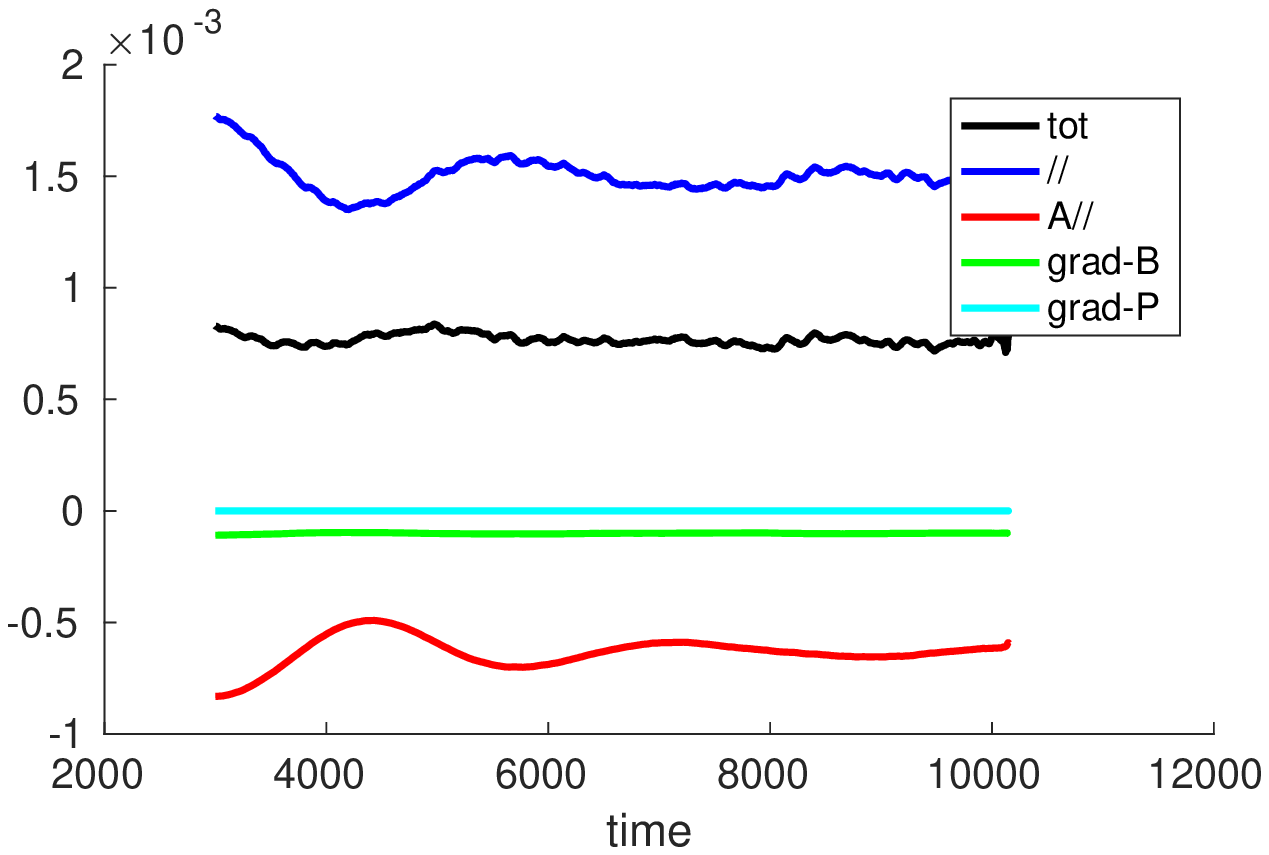}
\label{ITG_EM}
\\
c)&d)
\\
&
\\
\includegraphics[height=4cm,width=7cm]{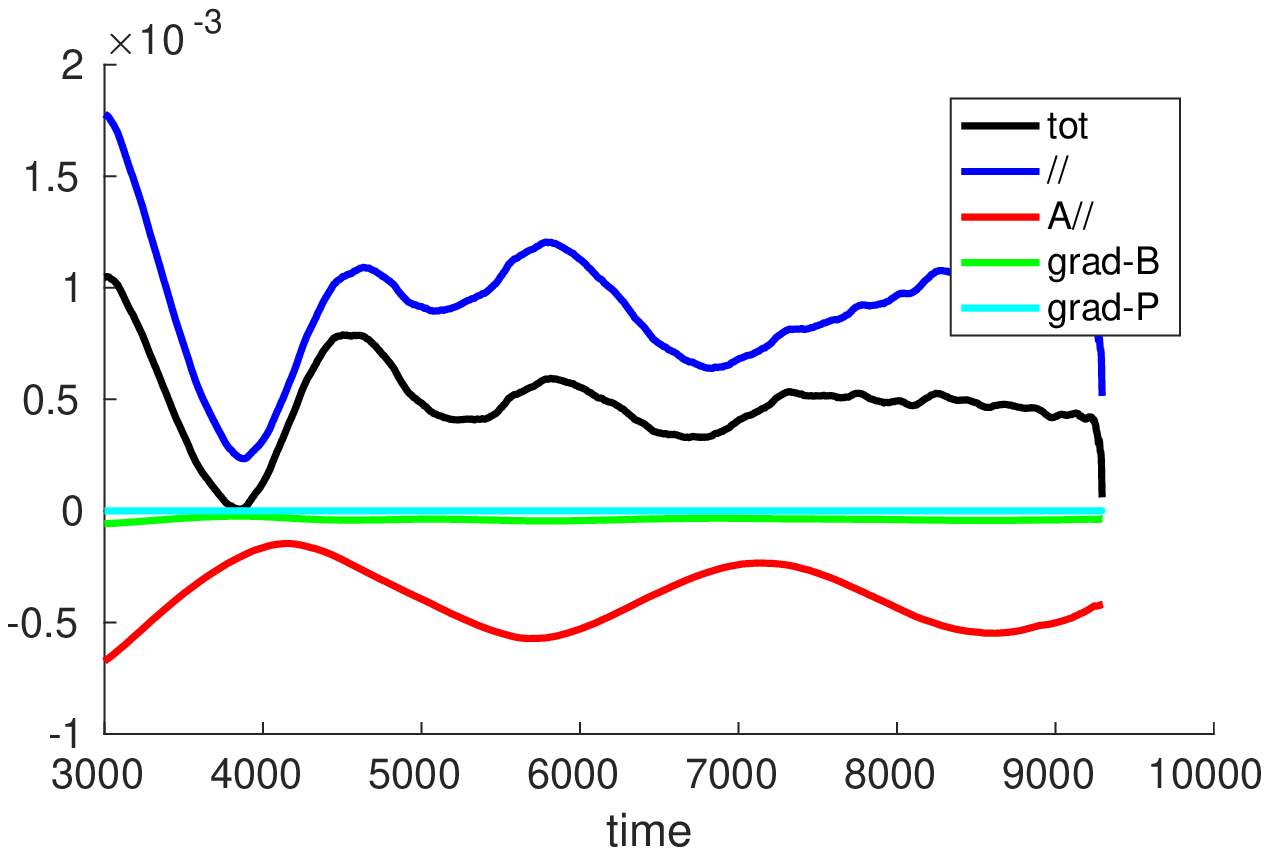}
\label{mode_change}&
\includegraphics[height=4cm,width=7cm]{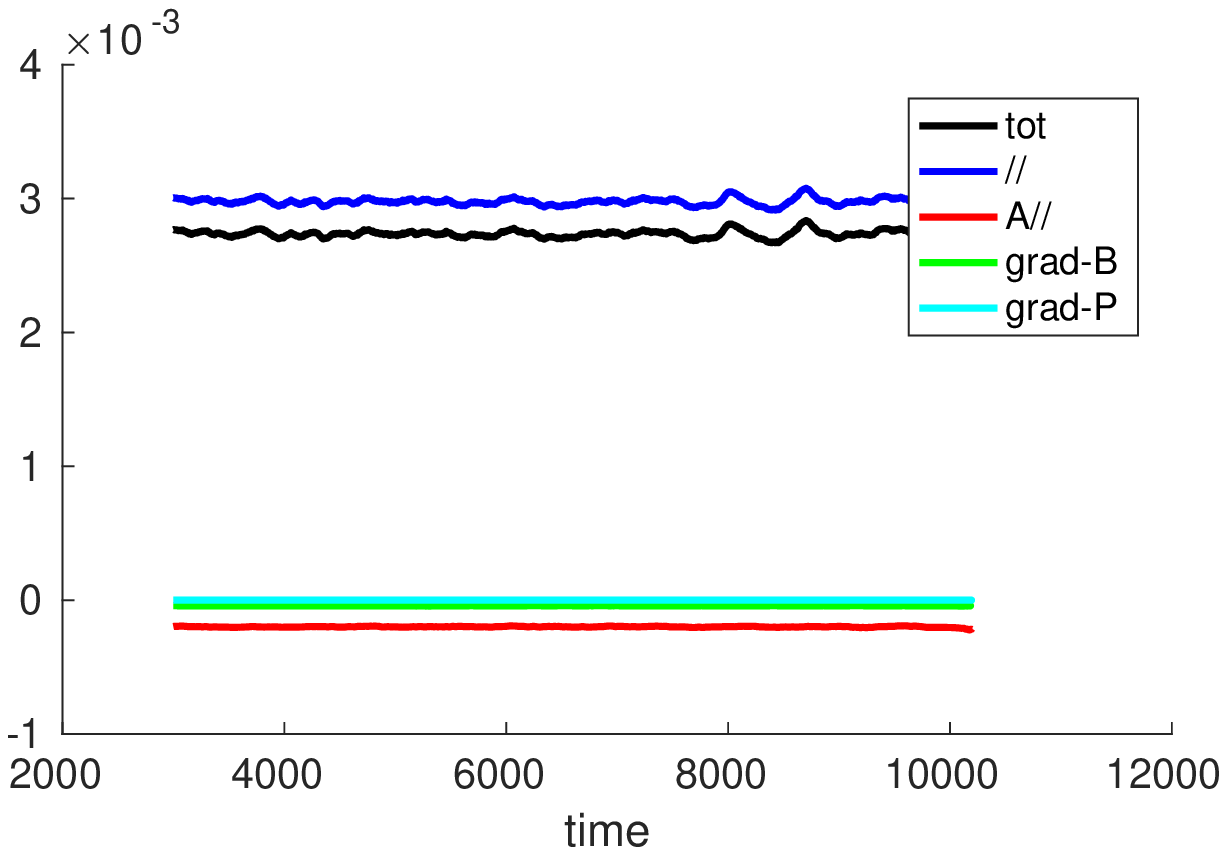}
\label{KBM}
\\
e)&f)
\\
&
\\
\end{tabular}
\caption{Time evolution of the contributions $\bm v_{\|,s}$, $\left\langle A_{\|}\right\rangle\bm v_{\mathrm{curv},s}$, $\bm v_{{\bm\nabla B},s}$ to the $\dd \mathcal{E}_F/\dd t$ diagnostics given by Eq.~(\ref{dEf_dt}), corresponding to the linear CYCLONE base case numerical configuration. The black curve corresponds to the total value of $\dd\mathcal{E}_F/\dd t$. The blue curve gives the value associated with $\bm{v}_{\|,s}$. The red curve is the part of the curvature drift coupled with purely electromagnetic potential, $\left\langle A_{\|}\right\rangle \bm{v}_{\mathrm{curv},s}$. The green curve represents the curvature drift, $\bm{v}_{{\bm{\nabla}B},s}$. The cyan curve is set equal to zero since $\mathbf B\times \left(\bm\nabla\times{\mathbf B}\right)= -\bm\nabla p$ in a single fluid MHD equilibrium. }
\end{center}
\label{fig:general}
\end{figure}
 In Fig.~1, the $\dd \mathcal{E}_F/\dd t$ diagnostics is presented for different values of $\beta$. Figure 1a) represents the electrostatic ITG with $\beta=0.00001\%$, $\Delta E_F=0.99$ which is mainly destabilized by curvature contribution with $\bm{v}_{\bm\nabla B,s}>0$. Figure 1b):  $\beta$ is increased up to $0.0025\%$, $\Delta E_F=0.79$, which activates an additional destabilizing mechanism with $\left\langle A_{1\|}\right\rangle\bm{v}_{\rm{curv},s}>0$. Figure 1c): $\beta=0.06\%$ corresponds to $\Delta E_{F}=0.03$, i.e., the amount of electrostatic energy is almost equal to the amount of electromagnetic energy in the system: This corresponds to the exchange of the stabilizing and destabilizing mechanisms, i.e., the mode is now destabilized by kinetic effects with $\bm{v}_{\|,s}>0$. However, the growth rate and the frequency of the mode exhibit no bifurcation. Figure~1d): $\beta=0.5\%$ corresponds to the electromagnetic ITG mode with $\Delta E_F=-86.20$, destabilized by kinetic effects $\bm{v}_{\|,s}>0$. Figure~1e): $\beta=0.69\%$ with $\Delta E_F=-296.76$, a minimal value for the normalized field energy corresponds to the  ITG to KBM bifurcation in frequencies and the growth rates. Figure~1f): $\beta=1.125\%$ $\Delta E_F=-204.89$ corresponds to the high frequency KBM mode with $\bm{v}_{\|,s}>0$.
 
 In Fig.~2, we perform the $\beta$-scan with both energy conservation based diagnostics,  the power balance diagnostics given by Eq.~(\ref{gamma}) and the  $\Delta{\mathcal E}_F$ diagnostics. In Fig.~2a), the growth rate of the instabilities as a function of $\beta$ is calculated according to the power balance diagnostics given by Eq.~(\ref{gamma}). The minimum corresponds to the bifurcation from ITG to KBM. On Fig.~2b), the corresponding frequencies are presented: the transition between the slow (ITG) and the fast mode (KBM) happens at the value $\beta=0.65\%$, which corresponds to the minimum of the power balance diagnostics. Figures~2c) and 2d) represent the $\Delta{\mathcal{E}}_F$ diagnostics. The functional dependency on $\beta$ is investigated:  On Fig.~2c), the zero of the $\Delta{\mathcal E}_F(\beta)$ function corresponds to the transition from the electrostatic ITG to the electromagnetic ITG, which corresponds to the exchange of stabilizing/destabilizing mechanisms, from the curvature drift $\bm v_{\rm curv,s}$ to the kinetic mechanisms driven by $\bm{v}_{\|,s}$. The minimum of the $\Delta{\mathcal E}_F(\beta)$ function corresponds to the bifurcation between the ITG and KBM modes, i.e., the minimum of the power balance diagnostics from Fig. 2a). Figure~2d) represents the part of Fig.~2c), with low $\beta$, featuring zero of $\Delta\mathcal{E}_F(\beta)$ function.
 
\begin{table}[htp]
\begin{center}
\begin{tabular}{|l l l l|}
\hline
&&&\\
$\beta$ \hspace{1cm} & $\omega_{\mathrm{Orb5}}$ \hspace{1cm}& $\gamma_{\mathrm{Orb5}}$\hspace{1cm} & $\Delta E_F$ 
\\ in $\%$ \hspace{1cm}            &  in $c_s/R_0$   \hspace{1cm}     &   in $c_s/R_0$ \hspace{1cm}        & \\ 
&&&\\
\hline
&&&\\
$10^{-5}$             &$0.667$               & $0.533$                    & $0.999$\\
$10^{-4}$             &$0.661$               & $0.527$                    &$0.999$ \\
%$0.025$               &$0.665$               & $0.540$                    &$0.794$ \\
$0.025$               &$0.665$               & $0.540$                    &$0.794$ \\
$0.05$                &$0.613$                & $0.544$                    &$0.313$ \\
$0.055$               &$0.613$               & $0.544$                    &$0.182$ \\
$0.06$                 &$0.597$               & $0.511$                    &$0.038$ \\
$0.065$               &$0.630$               & $0.486$                    &$-0.117$ \\
$0.068$               &$0.620$               & $0.555$                    &$-0.217$ \\
$0.07$                 &$0.677$               & $0.516$                    &$-0.284$ \\
$0.075$               &$0.677$               & $0.516$                    &$-0.4649$ \\
$0.1$                   &$0.651$               & $0.497$                    &$-1.576$ \\
$0.15$                 &$0.710$               & $0.477$                    &$-4.967$ \\
$0.25$                 &$0.773$               & $0.464$                    &$-17.870$ \\
$0.45$                 &$0.741$               & $0.391$                    & $-86.200$\\
$0.5$                 &$0.850$               & $0.345$                    & $-117.734$\\
$0.6$                   &$0.910$               & $0.270$                    &$-205.114$ \\
$0.65$                 &$1.260$               & $0.225$                    &$-256.218$ \\
$0.68$                 &$3.255 $               & $0.079$                    &$-283.435$ \\
$0.689$               &$3.189$              & $0.064$                    &$-291.325$ \\
$0.69$                 &$3.456$              & $0.0067$                    &$-293.262$ \\
$0.7$                   &$3.205$               & $0.152 $                    &$-292.720$ \\
$0.75$                 &$2.870$               & $0.196$                    &$-197.360$ \\
$0.8$                   &$2.701$               & $0.438$                    &$-202.060$ \\
$0.9$                   &$2.516$               & $0.809$                    &$-208.500$ \\
$1.000$               &$2.365$               & $1.083$                     &$-206.237$ \\
$1.125$               &$2.171$               & $1.337$                     &$-204.897$ \\
$1.25$                 &$2.191$               & $1.532$                     &$-205.668$ \\
&&&\\
\hline
\end{tabular}
\end{center}
\caption{\label{tab3}Data corresponding to analysis of linear electromagnetic simulations provided on Fig.~2.}
\end{table}%

\begin{figure}[h!]
\begin{center}
\begin{tabular}{c c}
\includegraphics[height=4cm,width=7cm]{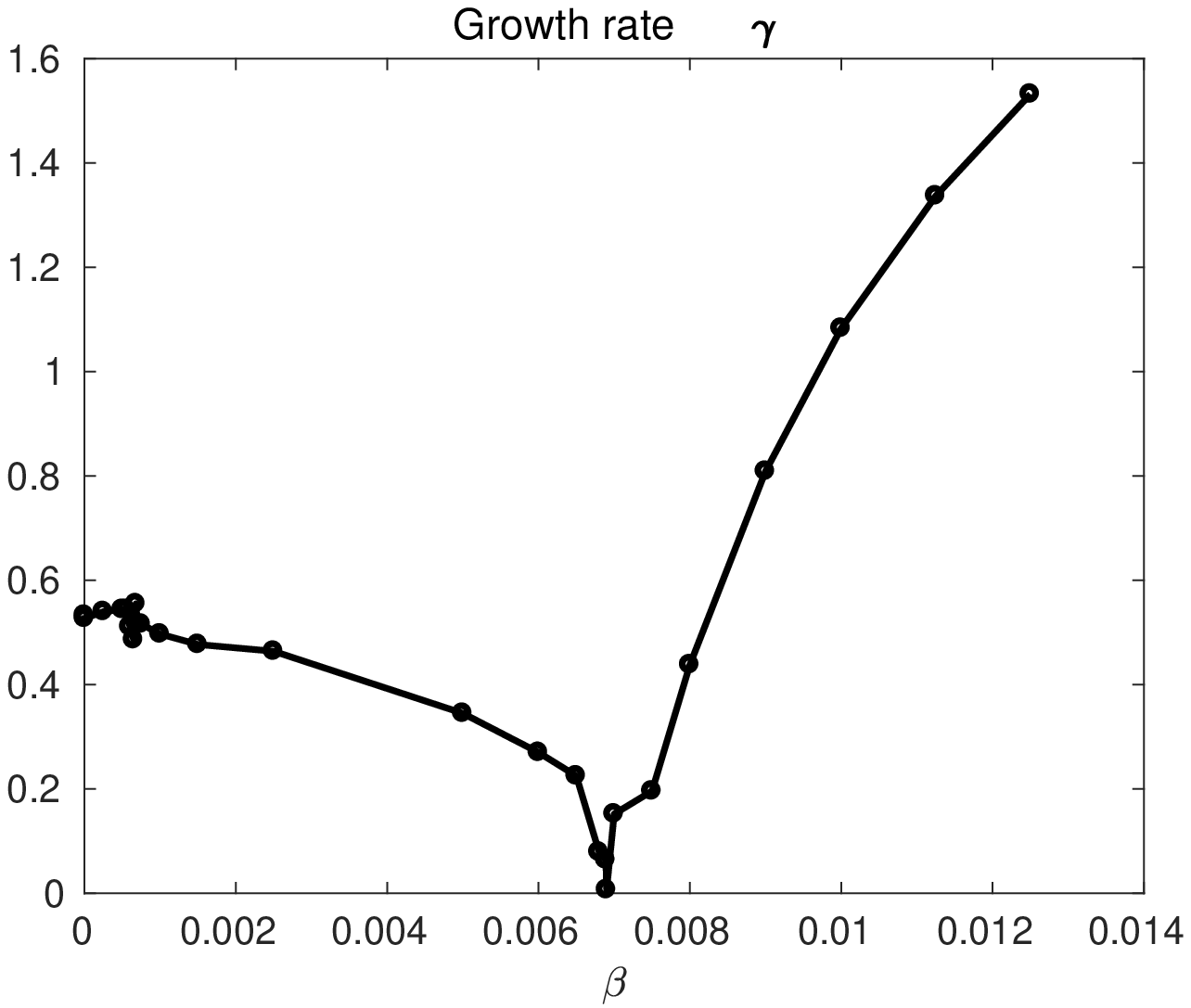}
&
\includegraphics[height=4cm,width=7cm]{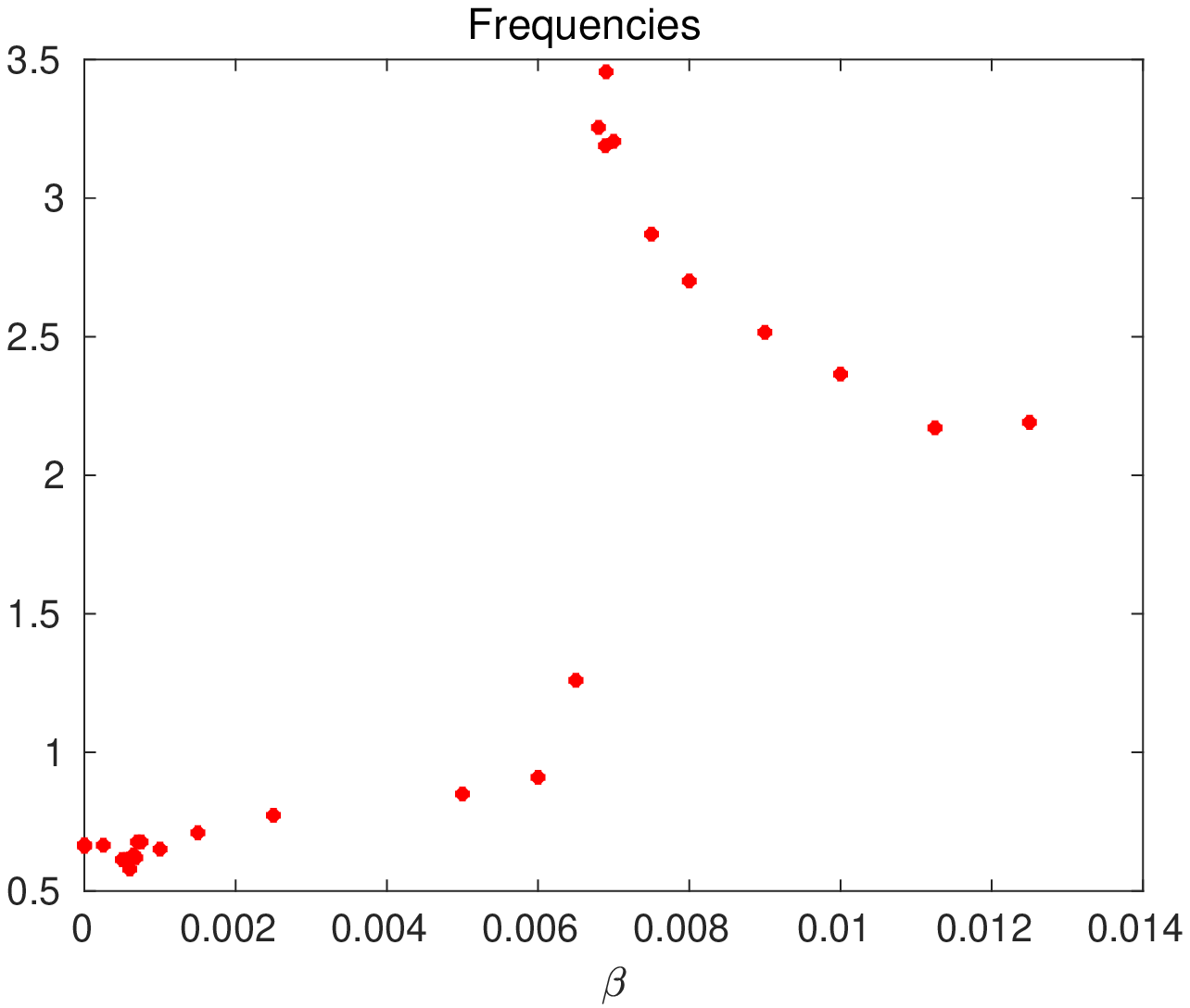}
\\
a) & b)
\\
&
\\
\includegraphics[height=4cm,width=7cm]{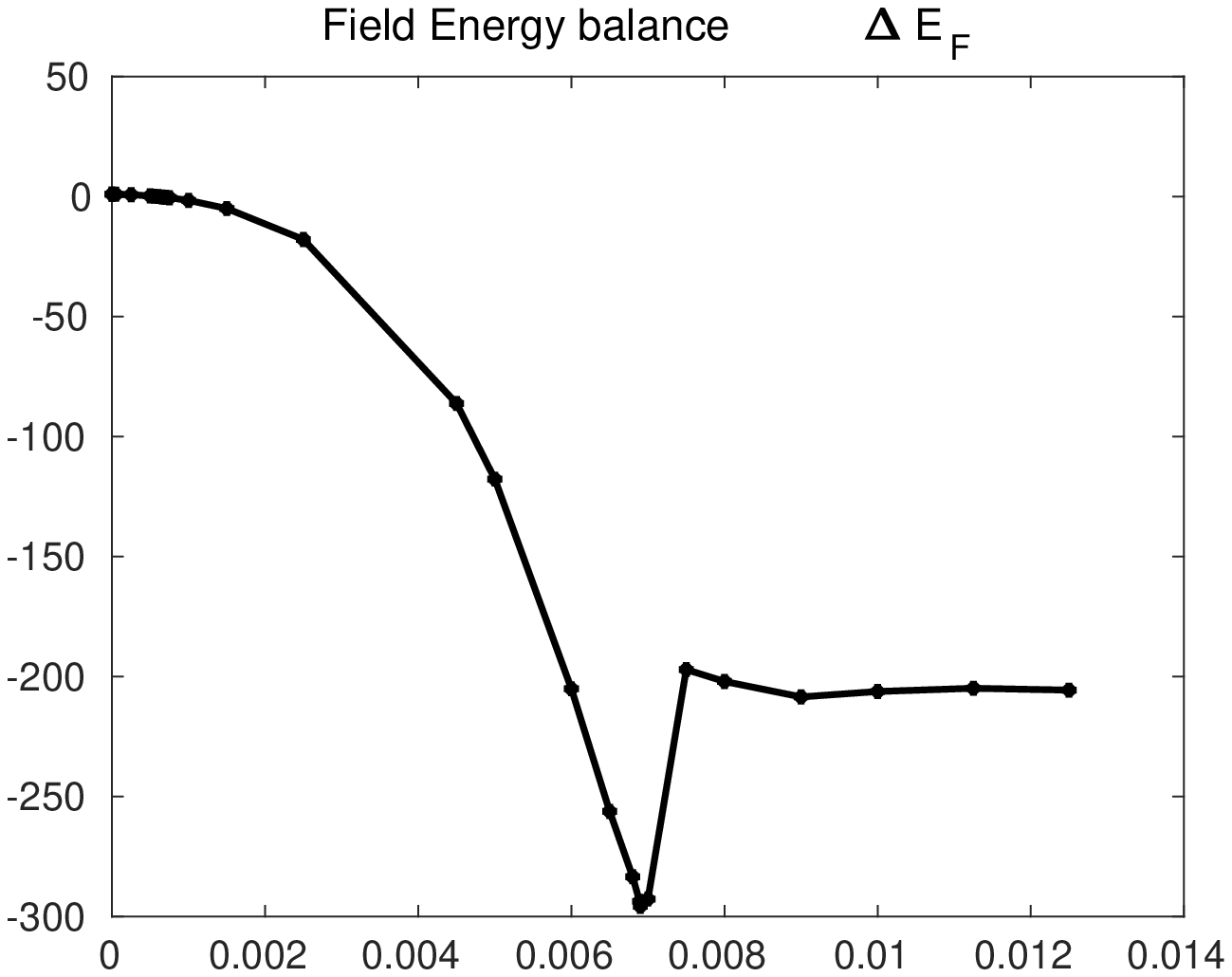}
&
\includegraphics[height=4cm,width=7cm]{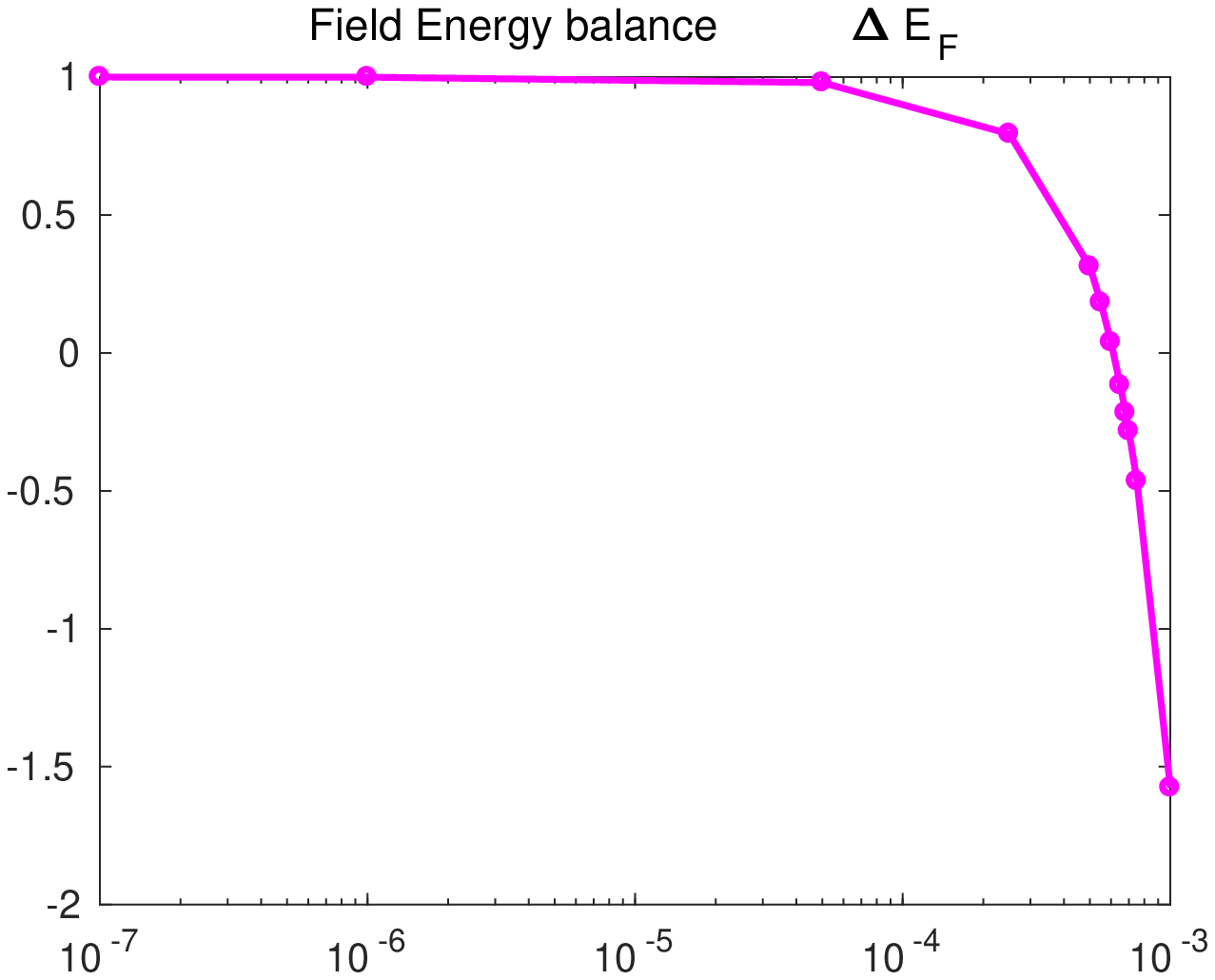}
\\
c) & d)
\label{KBM}
\end{tabular}
\caption{Power balance diagnostics, Figs.~3a) and~3b), and the $\Delta{\mathcal E}_F$ diagnostic, Figs.~3c) and 3d), as functions of $\beta$. }
\end{center}
\end{figure}

%%%%%%
\section{\label{sec:conclusion} Conclusions}
%%%%%%

In this work, the energy exchange channels leading to destabilisation of the electromagnetic instabilities in global gyrokinetic simulations with \orb{} code have been identified.  First, it has been observed through the simulations that with increasing $\beta$ the contribution of the magnetic curvature in the ITG mode destabilisation decreases together with the mode growth rate. It confirms the results of previous studies, indicating the stabilizing role of the magnetic $\beta$ on the ITG instability.
Second, the implementation of the diagnostics, issued from Noether's method, allows one to investigate the transition from the electrostatic towards the electromagnetic ITG regime. 
 Following the contributions to the energy time variation, the transition from the destabilizing role of curvature drift towards its stabilizing role is identified.
 The implementation of the energy invariant based diagnostic $\Delta{\mathcal E}_F$ allowed to systematically analyse essential features of the electromagnetic instabilities, for instance, by looking at the zeros of $\Delta{\mathcal E}_F$  corresponding to the ITG instability mechanism exchange and the minimum of that function corresponding to the bifurcation between the ITG and the KBM mode.
We observe that the transition of the instabilities generating mechanisms in the framework of the gyrokinetic theory follows the results obtained in previous studies realised in the framework of fluid and kinetic approaches \cite{Horton_book}. This transition happens at $\beta=0.06\%$, which follows previous theoretical predictions for $\beta> m_e/m_i\sim0.055\%$. Finally, it has been identified that the mode bifurcation towards the kinetic ballooning mode dominated regime happens at $\beta=0.65\%$, i.e., at a $\beta$ value $10$ times higher  than  the $\beta$ value at which mixing between the ITG (slow) and KBM (fast) modes occurs together with the destabilizing mechanisms exchange.

%%%%%%
\section*{\label{sec:acknowledgement} Acknowledgements}
%%%%%%
The first author thanks P.J.~Morisson and J.D. Meiss for useful advices and discussions.
This material is based upon work supported by the National Science Foundation under Grant No.~DMS-1440140 while NT and CC were in residence at the Mathematical Sciences Research Institute in Berkeley, California, during the Fall 2018 semester.
This work has been carried out within the framework of the French Federation for Magnetic Fusion Studies (FR-FCM) and of the Eurofusion consortium, and has received funding from the Euratom research and training programme 2014-2018 and 2019-2020 under grant agreement No.~ 633053. The views and opinions expressed herein do not necessarily reflect those of the European Commission. 

%%%%%%
%\section*{References}
%%%%%%

%\bibliographystyle{unsrt}
%\bibliography{PPCF_2018}

\begin{thebibliography}{30}

\bibitem{Wesson}
J.~A. {Wesson}.
\newblock {\em {Tokamaks}}.
\newblock {Oxford Science Publication, 4 th Edition}, 2011.

\bibitem{Horton_book}
W.~{Horton}.
\newblock {\em {Turbulent Transport in Magnetized Plasmas}}.
\newblock {World scientific}, 2012.

\bibitem{Rewoldt_1998}
G.~Rewoldt, M.~A. Beer, M.~S. Chance, T.~S. Hahm, and Z.~Lin.
\newblock Sheared rotation effects on kinetic stability in enhanced confinement
  tokamak plasmas, and nonlinear dynamics of fluctuations and flows in
  axisymmetric plasmas.
\newblock {\em Physics of Plasmas}, 5(5):1815--1821, 1998.

\bibitem{Hong_a_1989}
B.~G. Hong, W.~Horton, and D.~I. Choi.
\newblock Drift-alfven kinetic stability theory in the ballooning
  mode-approximmation.
\newblock {\em Physics of Fluids. B, Plasma Physics}, 1(8):1589--1599, 1989.

\bibitem{Hong_b_1989}
B.~G. Hong, W.~Horton, and D.~I. Choi.
\newblock Pressure gradient-driven modes in finite beta toroidal plasmas.
\newblock {\em Plasma physics and controlled fusion}, 31(8):1291--1303, 1989.

\bibitem{Kim_1993}
J.~Y. Kim, W.~Horton, and J.~Q. Dong.
\newblock Electromagnetic effect on the toroidal ion temperature-gradient mode.
\newblock {\em Physics of Fluids. B, Plasma Physics}, 5(11):4030--4039, 1993.

\bibitem{Brizard_Hahm}
A.~J. {Brizard} and T.~S. {Hahm}.
\newblock {Foundations of nonlinear gyrokinetic theory}.
\newblock {\em Reviews of Modern Physics}, 79:421, 2007.

\bibitem{Tronko_Chandre_2018}
N.~{Tronko} and C.~{Chandre}.
\newblock {Second order Gyrokinetic theory: from particle to gyrocenter}.
\newblock {\em {Journal of Plasma Physics}}, 84:925840301, 2018.

\bibitem{Garbet_Idomura_2010}
X.~{Garbet}, Y.~{Idomura}, L.~{Villard}, and T.~H. {Watanabe}.
\newblock {Gyrokinetic simulations of turbulent transport}.
\newblock {\em {Nuclear Fusion}}, 50:043002, 2010.

\bibitem{Jolliet_2007}
S.~{Jolliet}, A.~{Bottino}, P.~{Angelino}, R.~{Hatzky}, T.~M. {Tran}, B.~F.
  {Mcmillan}, O.~{Sauter}, K.~{Appert}, Y.~{Idomura}, and L.~{Villard}.
\newblock {A global collisionless PIC code in magnetic coordinates}.
\newblock {\em Computer Physics Communications}, 177:409, 2007.

\bibitem{Jenko_2000}
F.~{Jenko}, W.~{Dorland}, M.~{Kotschenreuther}, and B.~N. {Rogers}.
\newblock {Electron temperature gradient driven turbulence}.
\newblock {\em {Physics of plasmas}}, 7(5):1904--1910, 2000.

\bibitem{Bottino_Sonnendruecker}
A.~{Bottino} and E.~{Sonnendr\"ucker}.
\newblock {Monte Carlo Particle-In-Cell methods for the simulation of the
  Vlasov-Maxwell gyrokinetic equations}.
\newblock {\em Journal of Plasma Physics}, 81(5):435810501, 2015.

\bibitem{Goerler_2011}
T.~{Goerler}, X.~{Lapillonne}, S.~{Brunner}, {Dannert} T., F.~{Jenko},
  F.~{Merz}, and D.~{Told}.
\newblock {The global version of the gyrokinetic turbulence code GENE }.
\newblock {\em {Journal of Computational Physics}}, 230(18):7053 -- 7071, 2011.

\bibitem{Lanti_2018}
E.~Lanti, B.~McMilan, N.~Tronko, and L.~Villard.
\newblock \orb{}: a global electromagnetic gyrokinetic code using the pic
  approach in toroidal geometry.
\newblock {\em in preparation}, 2018.

\bibitem{Goerler_Tronko_2016}
T.~{Goerler}, N.~{Tronko}, W.~A. {Hornsby}, A.~{Bottino}, R.~{Kleiber},
  C.~{Norcini}, V.~{Grandgirard}, F.~Jenko, and E.~{Sonnendr\"ucker}.
\newblock {Intercode comparison of gyrokinetic global electromagnetic modes}.
\newblock {\em {Physics of Plasmas}}, 23:072503, 2016.

\bibitem{TBS_2016}
N.~{Tronko}, A.~{Bottino}, and E.~{Sonnendr\"ucker}.
\newblock {Second order gyrokinetic theory for {P}article{-}{I}n{-}{C}ell
  codes}.
\newblock {\em {Physics of Plasmas}}, 23:082505, 2016.

\bibitem{brizard_prl_2000}
A.~J. {Brizard}.
\newblock {New Variational Principle for the {V}lasov{-}{M}axwell Equations}.
\newblock {\em Physical Review Letters}, 84:5768, 2000.

\bibitem{Sugama_2000}
H.~{Sugama}.
\newblock {Gyrokinetic field theory}.
\newblock {\em Physics of Plasmas}, 7:466, 2000.

\bibitem{Cary_Brizard}
J.~R. {Cary} and A.~J. {Brizard}.
\newblock {Hamiltonian theory of guiding center motion}.
\newblock {\em Reviews of Modern Physics}, 81:693, 2009.

\bibitem{Dimits_2000}
A.~{Dimits}, G.~{Bateman}, M.A. {Beer}, B.I. {Cohen}, W.~{Dorland}, G.W.
  {Hammett}, C.~{Kim}, J.E. {Kinsley}, M.~{Kotschenreuter}, A.H. {Kritz}, L.L.
  {Lao}, J.~{Mandrekas}, W.M. {Nevins}, S.E. {Parker}, A.J. {Redd}, D.E.
  {Schumaker}, R.~{Sydora}, and J.~{Weiland}.
\newblock {Comparison and physics basis of tokamak transport models and
  turbulence simulations}.
\newblock {\em Physics of Plasmas}, 7(3):969--983, 2000.

\bibitem{Greenfield_1997}
C.~M. Greenfield, J.~C. DeBoo, T.~H. Osborne, F.~W. Perkins, and M.~N.
  Rosenbluth.
\newblock Enhanced fusion performance due to plasma shape modification of
  simulated {ITER} discharges in {DIII-D}.
\newblock {\em Nuclear fusion}, 37(9):1215--1228, 1997.

\end{thebibliography}

\end{document}